\documentclass[]{raa-rad}                  
\usepackage{graphicx,times}             
\usepackage{color}
\usepackage{fancybox}

\newcommand{\boldsymbol}[1]{\mbox{\boldmath $#1$}}

\newcommand{\degree}{\hbox{$^\circ$}}

\begin{document}

\title{Bipolar Outflows as a Repulsive Gravitational Phenomenon \\{\large -- Azimuthally Symmetric Theory \textit{of} Gravitation (II) }}

   \volnopage{{\bf 2010} Vol.\ {\bf 10} No. {\bf XX}, 000--000}
   \setcounter{page}{1}

   \author{Golden Gadzirayi Nyambuya\footnotetext{$^{*}$Supported by the Republic \textit{of} South Africa's  National Research Foundation and the North West University, and Germany's 
DAAD Programme \textit{via} the University \textit{of} K$\ddot{\rm{o}}$ln.}
}

\institute{North West University (Potchefstroom Campus), School of
Physics (Unit for Space Research), Private Bag X6001, Potchefstroom
$2531$, Republic of South Africa, Email: {\it gadzirai@gmail.com}\\
\vs \no
   {\small Received 2010 May 14; accepted 2010 June 27 }
}

\abstract{This  reading is part in a series on the  Azimuthally Symmetric Theory \textit{of} Gravitation (ASTG) set-out in Nyambuya 
(\cite{nyambuya10a}). This theory is built on Laplace-Poisson's well known equation and it has been shown therein (Nyambuya \cite{nyambuya10a}) that 
the ASTG  is capable of explaining -- from a purely classical physics standpoint; the precession of the perihelion of solar planets as being a 
consequence of the azimuthal symmetry emerging from the spin of the Sun. This symmetry has and must have an influence on the emergent gravitational 
field.  We show herein that the emergent equations from the ASTG -- under some critical conditions determined by the spin -- do possess repulsive 
gravitational fields in the polar regions of the gravitating body in question. This places the ASTG on an interesting pedal  to infer the origins of 
outflows as a repulsive gravitational phenomena. Outflows are an ubiquitous phenomena found in star forming systems and their true origins is a question 
yet to be settled. Given the current thinking on their origins, the direction that the present reading takes is nothing short of an asymptotic break 
from conventional wisdom; at the very least, it is a complete paradigm shift as gravitation is not at all associated; let alone considered to have 
anything to do with the out-pour of matter but is thought to be an all-attractive force that tries only to squash matter together into a single point. 
Additionally, we show that the emergent Azimuthally Symmetric Gravitational Field  from the ASTG strongly suggests a solution to the supposed Radiation 
Problem that is thought to be faced by massive stars in their process of formation. That is, at $\sim 8-10\,\mathcal{M}_{\odot}$, radiation from the 
nascent star is expected to halt the accretion of matter onto the nascent star. We show that in-falling material will fall onto the equatorial disk and 
from there, this material will be channelled onto the forming star \textit{via} the equatorial plane thus accretion of mass continues well past the 
curtain value of $\sim 8-10\,\mathcal{M}_{\odot}$ albeit \textit{via} the disk. Along the equatorial plane, the net force (with the radiation force 
included) on any material there-on right up-till the surface of the star, is directed toward the forming star, hence accretion of mass by the nascent 
star is un-hampered. 
\\
\textbf{PACS (2010):} $97.10.$Bt, $97.10.$Gz, $97.10.$Fy
}

   \authorrunning{G. G. Nyambuya}            
   \titlerunning{Bipolar Outflows as a Repulsive Gravitational Phenomenon}  

   \maketitle

%

{\renewcommand{\theequation}{\textcolor{blue}{\arabic{equation}}}
 \renewcommand{\thefigure}{(\textcolor{red}{\arabic{figure}})}
 \renewcommand{\thetable}{(\textcolor{red}{\Roman{table}})}
 
%
\section{Introduction}

Champagne like bipolar molecular outflows are an unexpected natural phenomenon that grace the star formation podium. Bipolar molecular outflows are the 
most spectacular physical phenomenon intimately associated with newly formed stars. Studies of bipolar outflows reveal that they [bipolar outflows] are 
ubiquitous toward High Mass Star (HMS) forming regions. These outflows in HMS forming regions are far more massive and energetic than those found 
associated with Low Mass Stars (LMS) forming regions (see \textit{e.g.} Shepherd \textit{\&} Churchwell \cite{shepherd96a}; Shepherd \textit{\&} 
Churchwell \cite{shepherd96b}; Zhang \textit{et al.} \cite{zhang01}; Zhang \textit{et al.} \cite{zhang05}; Beuther \cite{beuther02}). Obviously, this 
points to a correlation between the mass of the star and the outflow itself.  Independent studies have established the existence of such a correlation. 
The mass outflow rate $\dot{\mathcal{M}}_{out}$ has been shown to be related to the bolometric luminosity $\mathcal{L}$  by the relationship: 
$\dot{\mathcal{M}}_{out}\propto \mathcal{L}^{0.60}_{star}$, and this is for stars in the luminosity range: 
${0.30}\mathcal{L}_{\odot}\leq \mathcal{L}_{star} \leq {10}^{5}\mathcal{L}_{\odot}$ (we shall use the term luminosity to mean bolometric luminosity). 
Another curious property of outflows is that the mass-flow rate, $\dot{\mathcal{M}}_{out}$, is related to the speed of the molecular outflow 
$\dot{\mathcal{M}}_{out}\propto V^{-\gamma}_{out}$ where $\gamma\sim {1.80}$ and $V_{out}$ is the speed of the outflow. How and why outflows come to 
exhibit these properties is an interesting field of research that is not part of the present reading. However, we shall show that these relationships 
do emerge from our proposed ASTG Outflow Model. In the present, we simple want to show that an outflow model emerges from the ASTG model. We set herein 
the mathematical foundations for such a model. Once we have a fully-fledged mathematical model, we shall move on to building a numerical model 
(\textit{i.e.} computer code). Once this computer code is available, an endeavor to answer the above and other questions surrounding the nature of 
outflows will be made.

Pertaining to their association with star formation activity, it is believed that molecular outflows are a necessary part of the star formation process 
because their existence may explain the apparent angular momentum imbalance. It is well known that the amount of initial angular momentum in a typical 
star-forming molecular cloud core is several orders of magnitude too large to account for the observed angular momentum found in formed or forming 
stars (see \textit{e.g.} Larson \cite{larson03b}). The sacrosanct Law \textit{of} Conservation of angular momentum informs us that this angular 
momentum can not just disappear into the oblivion of interstellar spacetime. So, the question is where does this angular momentum go to? It is here that 
outflows are thought to come to the rescue as they can act as a possible agent that carries away the excess angular momentum. This angular moment, if it
 where to remain as part of the nascent star, it would, \textit{via} the  strong centrifugal forces, tear the star apart. This however does not explain, 
why they exist and how they come to exist but simply posits them as a vehicle needed to explain the mystery of ``{The Missing Angular Momentum Problem}''
 in star forming systems and the existence of stars in their intact and compact form as stable firery balls of gas.

In the existing literature, \textit{viz}  the question why and how molecular outflows exist, there are about four proposed leading models that endeavor 
to explain the aforesaid. These four major proposals are:\\

\noindent \textbf{\underline{Wind Driven Outflow Model}:} In this model, a wide-angle radial wind blows into the stratified surrounding ambient material,
 forming a thin swept-up shell that can be identified as the outflow shell (see Shu \textit{et al.} \cite{shu91}; Li \textit{\&} Shu \cite{li96}; Matzner
 \textit{\&} McKee \cite{matzner99}).\\

\noindent \textbf{\underline{Jet Driven Bow Shocks Model}:} In this model, a highly collimated jet propagates into the surrounding ambient material 
producing a thin outflow shell around the jet (see Raga \textit{et al.} \cite{raga93a}; Masson \textit{\&} Chernin \cite{masson93}).\\

\noindent \textbf{\underline{Jet Driven Turbulent Outflow Model}:} In this model, Kelvin-Helmholtz instabilities along the jet and or environmental 
boundary leading to the formation of a turbulent viscous mixing layer, through which the molecular cloud gas in entrained (see Cant\'o \textit{\&} 
Raga \cite{canto91}; Raga \textit{et al.} \cite{raga93b}; Stahler \cite{stahler94}; Lizano \textit{\&} Giovanardi \cite{lizano95}; 
Cant\'o \textit{et al.} \cite{canto03}).\\

\noindent\textbf{\underline{Circulation Flows Model}:} In this model, the molecular outflow is not entrained by an underlying wind jet but is rather 
formed by in-falling matter that is deflected away from the protostar in the central torus of high magneto-hydrodynamic pressure through a quadrupolar 
circulation pattern around the protostar and is accelerated above escape speeds by local heating (see Fiege \textit{\&} Henriksen \cite{fiege96a}; 
Fiege \textit{\&} Henriksen \cite{fiege96a}).\\

\noindent All these \textit{ad hoc} models and some that are not mentioned here explain outflows as a feedback effect. The endeavor of the work 
presented in this reading is to make an alternative suggestion albeit a complete, if not a radical departure from the already existing models briefly 
mentioned above. Our model flows naturally from the Laplace-Poison equation, namely from the Azimuthally Symmetric Theory of Gravitation (ASTG) laid 
down in Nyambuya (\cite{nyambuya10a}) (hereafter Paper I). This model is new and has never before appeared in the literature. Because we are at the 
stage of setting this model, we see no need to get into the details of the existing models as this would lead to an unnecessary digression, confusion, 
and an un-called for lengthy reading.

Our model is a complete departure from the already existing models because, of all the agents that could lead to outflows, gravitation is not even 
considered to be a possible agent because it is thought of as, or assumed to be, an all-attractive force. Actually, the idea of a gravitating body such 
as a star producing a repulsive gravitational field, is at the very least unthinkable. Contrary to this, we show here that an azimuthally symmetric 
gravitational system does \textit{in-principle} give rise to a bipolar repulsive gravitational field and this -- in our view, clearly suggests that 
these regions of repulsive gravitation, possibly are the actual driving force of the bipolar molecular outflows. We also see that the ASTG provides a  
neat solution (possibly and very strongly so) to the so-called Radiation Problem  thought to bedevil and bewilder the formation of HMSs (see Larson 
\textit{\&} Starrfield \cite{larson71}; Kahn \cite{kahn74}; Bonnell \textit{et al.} \cite{bonnell98}; Bonnell \textit{\&} Bate \cite{bonnell02}; Palla 
\textit{\&} Stahler \cite{palla93}) and as-well the observed \textit{Ring of Masers} (Bartkiewicz \textit{et al.} \cite{bartkiewicz08}, 
\cite{bartkiewicz09}).

We need to reiterate this so as to make it clear to our reader, that, the work presented in this reading is meant to lay down the mathematical 
foundations of  the outflow  model emergent from the ASTG. It is not a comparative study of this outflow model  with those currently in existence. 
We believe we have to put thrust on lying down these ideas and only worry about their plausibility, \textit{i.e.} whether or not they correspond with 
experience and only thereafter make a literature wide comparative study. Given that this model flows naturally from a well accepted equation (the Poisson-Laplace equation), against the probability of all unlikelihood, this model should have a bearing with reality. If it does not have a bearing with reality, then, at the very least, it needs to be investigated since this solution of the Poisson-Laplace equation has not been explored anywhere in the literature\footnote{In our exhaustive survey of the accessible literature, we have not come across a treatment of the Poisson-Laplace equation as is done in the present, hence our proclamation that this solution of the Poisson-Laplace equation is the first such.}.

Also, we should say that as we build this model, we are doing this with expediency, that is, watchful of what experience dictates, at the end of the day,
 if our efforts are to bear any fruits, our model must correspond with reality. This literature wide comparative study is expected to be done once a 
mathematical model of our proposed outflow model is in full-swing. This mathematical model is expected to form part of the future works where only-after 
that,  it would make sense then to embark on this literature wide comparative study. How does one compare a baby human-being to a human-embryo? It does not make sense, does it? Should not the baby be born first and only thereafter a comparative study be conducted of this baby with those babies already in existence?  We hope the reader concurs with us that this is perhaps the best way to set into motion a new idea amid a plethora of ideas that champion a similar if not the same endeavor.

Further, we need to say this;  that, as already  stated above,  the direction that the present reading takes is nothing short of an asymptotic break 
from conventional wisdom; at the very least, it is a complete paradigm shift as gravitation is not at all associated; let alone considered to have 
anything to do with the out-pour of matter but is thought to be an all-attractive force that tries only to squash matter together into a single point. 
Because of this reason, that, the present is ``nothing short of an asymptotic break from conventional wisdom'' and that  ``at the very least, it is a 
complete paradigm shift'', we strongly believe that this is enough to warrant the reader's attention to this seemingly seminal theoretical discovery.

The synopsis of this reading is as follows. In the subsequent section, we present the theory to be used in setting up the proposed ASTG Outflow Model. 
In \S ($3$), we revisit the persistent problem of the ASTG model, that of ``The ASTG's Undetermined Parameter Problem''. Therein, we present what we 
believe may be a solution to this problem. As to what really these parameters may be, this is still an open question subject to debate. In \S ($4$), we 
present the main findings of the present reading, that is. the repulsive bipolar gravitational field and therein we argue that this field  fits the 
description of outflows. We present this for both the empty and non-empty space solutions of the Poisson-Laplace equation. In \S ($5$), we look at the 
anatomy of the outflow model, \textit{i.e.} the switching on and off outflows, the nature of the repulsive polar field, the emergent shock rings and the 
collimation factor of these outflows. In \S ($6$), we show that the ASTG model posits what strongly appears to be a perdurable solution to the so-called 
Radiation Problem that is thought to be faced by massive stars during their formation process. Lastly, in \S ($7$), we give a general discussion and 
make conclusion that cane be drawn from this reading.

Lastly, it is important that we mention here in the penultimate of this introductory section that this reading is fundamental in nature and because of 
this, we shall seek to begin whatever argument we seek to rise, from the soils of its very basic and fundamental level. This is done so that we are at 
the same level of understanding with the reader. With the aforesaid approach, if at any point we have errored, it would be easy to know and understand 
where and how we have errored.

\section{Theory\label{outf_theory}}

Newton's Law of universal gravitation can be written in a more general and condensed form as Poisson's Law, \textit{i.e.}:

\begin{equation}
\vec{\nabla}^{2}\Phi=4\pi G\rho,\label{poisson}
\end{equation}

\noindent where $\rho$ is the density of matter and $G=6.667\times10^{-11}\textrm{kg}^{-1}\textrm{ms}^{-2}$ is Newton's universal constant of 
gravitation and the operator $\vec{\nabla}^{2}$ written for a spherical coordinate system [see figure \ref{scoord} for the coordinate setup] is given 
by:

\begin{equation}
\vec{\nabla}^{2}= \frac{1}{r^{2}}\frac{\partial}{\partial r}\left(r^{2}\frac{\partial}{\partial r}\right)+\frac{1}{r^{2}\sin\theta}
\frac{\partial}{\partial \theta}\left(\sin\theta\frac{\partial}{\partial\theta}\right)+\frac{1}{r^{2}\sin^{2}\theta}\frac{\partial^{2}}{\partial 
\varphi^{2}},
\end{equation}

\noindent where the symbols have their usual meanings. For a spherically symmetric setting, the solution to Poisson's equation outside the vacuum 
space (where $\rho=0$) of a central gravitating body of mass $\mathcal{M}_{star}$  is given by the traditional inverse distance Newtonian gravitational 
potential which  is given by:

\begin{equation}
\Phi(r)=-\frac{G\mathcal{M}_{star}}{r},\label{npot}
\end{equation}

\noindent where $r$ is the radial distance from the center of the gravitating body. The Poisson equation for the case $(\rho=0)$ is known as the 
Laplace equation. The Poisson equation is an extension of the Laplace equation. Because of this, we shall generally refer to the Poisson equation as 
the Poisson-Laplace equation. In the case where there is material surrounding this central mass, that is $\mathcal{M}=\mathcal{M}(r)$, where:

\begin{equation}
\mathcal{M}(r)=\int^{r}_{0}\int^{2\pi}_{0}\int^{2\pi}_{0}r^{2}\rho(r,\theta,\varphi) \sin\theta  d\theta d\varphi dr,\label{mass}
\end{equation}

\noindent we must -- in (\ref{npot}), make the replacement: $\mathcal{M}_{star}\longmapsto\mathcal{M}(r)$. As already argued in Paper I, if the 
gravitating body in question is spinning, we ought to consider an Azimuthally Symmetric Gravitational Field (ASGF). Thus, we shall solve the 
azimuthally symmetric setting of (\ref{poisson}) for both cases of empty and non-empty space and show from these solutions that Poisson's equation 
entails a repulsive bipolar gravitational field. We shall assume that if one has the empty space solution, to obtain the non-empty space solutions, 
one has to make the replacement: $\mathcal{M}_{star}\longmapsto\mathcal{M}(r)$, just as is done in Newtonian gravitation. This is a leaf that we shall 
take from spherically symmetric Newtonian gravitation into the ASTG model.

\begin{figure}
\centering
\includegraphics[scale=1.0]{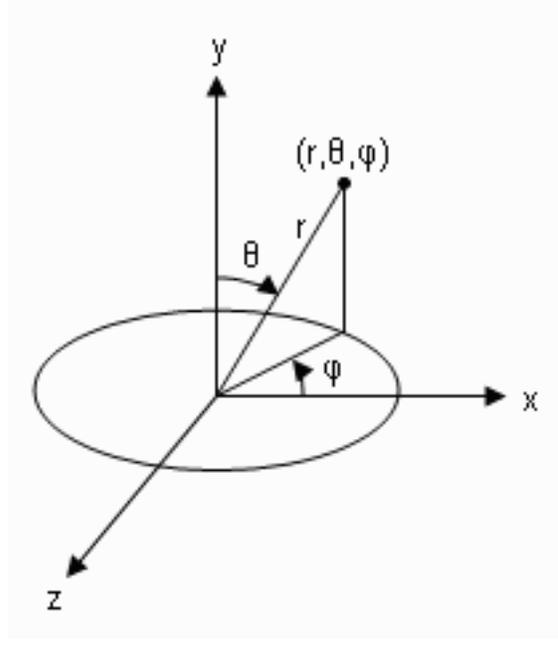}
\caption[\textbf{Generic Spherical Coordinate System}]{This figure shows a generic spherical coordinate system, with the radial coordinate denoted by 
$r$, the zenith (the angle from the North Pole; the co-latitude) denoted by $\theta$, and the azimuth (the angle in the equatorial plane; the longitude) 
by $\varphi$.}
\label{scoord}
\end{figure}

\subsection{Empty Space Solutions}

As already argued in Paper I, for a scenario or setting that exhibits azimuthal symmetry such as a spinning gravitating body as the Sun and also the 
stars that populate the heavens (where the unexpected and spectacular champagne like bipolar molecular outflows are the observed); we must have: 
$\Phi=\Phi(r,\theta)$. There-in Paper I, the Poisson equation for empty space has been ``solved'' for a spinning gravitating system  and the solution to 
it is:

\begin{equation}
\Phi(r,\theta)=-\sum^{\infty}_{\ell=0}\left[ \lambda_{\ell}c^{2}\left(\frac{G\mathcal{M}_{star}}{rc^{2}}\right)^{\ell+1}P_{\ell}(\cos\theta)\right],
\label{newpot}
\end{equation}

\noindent where $\lambda_{\ell}$ is an infinite set of dimensionless parameters with $\lambda_{0}=1$ and the rest of the parameters $\lambda_{\ell}$ for 
$(\ell>1)$, generally take values  different from unity. There-in Paper I, a suggestion as to what these parameters may be has been made. In \S ($3$) we 
go further and suggest a form for these parameters. This suggestion, if correct, puts the ASTG on a pedestal to make predictions without first seeking 
these values (\textit{i.e.} the $\lambda_{\ell}$'s) from observations.   We will show that there lays embedded in (\ref{newpot}) a solution that is such 
that the polar regions of the gravitating central body will exhibit a repulsive gravitational field. It is this repulsive gravitational field that we 
shall propose as the driving force causing the emergence of outflows. But, we must bare in mind that outflows are seen in regions in which the central 
gravitating body is found in the immensement of ambient circumstellar material, thus we must -- for the azimuthally symmetric case (where the central 
gravitating body is spinning), solve the Poisson-Laplace equation for the setting $(\rho\neq0)$. 

\subsection{Non-Empty Space Solutions}

Clearly, in the event that $(\rho\neq0)$ for the azimuthally symmetric case, we must have $\rho=\rho(r,\theta)$. In Paper~I, an argument has been 
advanced in support of this claim that: $\Phi(r,\theta)\Rightarrow\rho(r,\theta)$. Taking this as given, the question we wish to answer is; what form 
does $\Phi(r,\theta)$ take for a given mass distribution $\rho(r,\theta)$? or the reverse, what form does $\rho(r,\theta)$ take for a given  
$\Phi(r,\theta)$? It is reasonable and most logical to assume that the gravitational field is what influences the distribution of mass and not the other 
way round. Taking this as the case, then, we must have $\rho(r,\theta)=\rho(\Phi)$, \textit{i.e.} the distribution of the matter in any mass 
distribution must be a function of the gravitational field. We find that the form for $\rho(r,\theta)$ that meets the requirement: 
$\rho(r,\theta)=\rho(\Phi)$, and most importantly the requirement that to obtain the non-empty space solution from the empty space solution one simply 
makes the replacement: $\mathcal{M}_{star}\longmapsto\mathcal{M}(r)$, is:  

\begin{equation}
\rho(r,\theta)=-\frac{1}{4\pi G} \left[\frac{2}{r}\frac{\partial}{\partial r }-\frac{1}{r^{2}}\right]\frac{\partial \Phi(r,\theta)}{\partial\theta }
\label{rhochoice}.
\end{equation}

\noindent How did we arrive at this?  We have  to answer this question. To make life very easy for us to arrive at the answer, we shall write Poisson's 
equation in rectangular coordinates, \textit{i.e.}:

\begin{equation}
\left(\sum_{j=1}^{3}\frac{\partial^{2} }{\partial x^{2}_{j}}\right)\Phi(x,y,z)=4\pi G\rho(x,y,z)\label{pcomp1},
\end{equation}

\noindent where $x_{1}=x,x_{2}=y,x_{3}=z$. Now suppose we had a function $F(x,y,z)$ such that:

\begin{equation}
\left(\sum_{j=1}^{3}\frac{\partial }{\partial x_{j}}\right)^{2}F(x,y,z)=0.
\end{equation}

\noindent This equation can be written as:

\begin{equation}
\left(\sum_{j=1}^{3}\frac{\partial^{2} }{\partial x^{2}_{j}}\right)F(x,y,z)=-\left(\sum_{j}^{3}\sum_{i\neq j}^{3}
\frac{\partial^{2} }{\partial x_{i}\partial x_{j}}\right)F(x,y,z)\label{pcomp2}.
\end{equation}

\noindent Now, if and only if the gravitational potential did satisfy (\ref{pcomp1}), then, comparison of (\ref{pcomp1}) with (\ref{pcomp2}) requires 
the identification: $\Phi(x,y,z,)\equiv F(x,y,z)$, and as-well the identification:

\begin{equation}
\rho(x,y,z)=-\frac{1}{4\pi G}\left(\sum_{j}^{3}\sum_{i\neq j}^{3}\frac{\partial^{2} }{\partial x_{i}\partial x_{j}}\right)\Phi(x,y,z)\label{pcomp3}.
\end{equation}

\noindent What this means is that the non-linear terms of (\ref{pcomp1}) come about because of the presence of matter. Now, if we transform to 
spherical coordinates, it is now understood as to why and how we came to the choice of $\rho$ given in (\ref{rhochoice}). At the end of the day, what 
this means is that we can choose whatever form for $\Phi$, the density $\rho$ will have to conform and prefigure to this setting of the gravitational 
field \textit{via} (\ref{pcomp3}). Only and only after accepting (\ref{pcomp3}), do we have the mathematical legitimacy to choose to maintain the 
form (\ref{newpot}) which we found for the case of empty space such that in the place of $\mathcal{M}_{star}$ we now can put $\mathcal{M}(r)$, hence 
thus  in the case where a central gravitating condensation of mass is in the immensement of ambient circumstellar material,  we must have:

\begin{equation}
\Phi(r,\theta)=-\sum^{\infty}_{\ell=0}\lambda_{\ell}c^{2}\left(\frac{G\mathcal{M}(r)}{rc^{2}}\right)^{\ell+1}P_{\ell}(\cos\theta),\label{newpot2}
\end{equation}

\noindent where $\mathcal{M}(r)$ is given in (\ref{mass}). We believe this answers the question ``What form does $\rho(r,\theta)$ take for a given 
$\Phi(r,\theta)$?'' and at the same-time we have justified (\ref{rhochoice}) \textit{viz} how we have come to it. Importantly, it should be noted that  
the observed radial density profile is maintained by the choice (\ref{pcomp3}), \textit{i.e.} 
$\rho(r)=\int^{r}_{0}\int^{2\pi}_{0}r^{2}\rho(r,\theta)\sin\theta d\theta dr\propto r^{-\alpha_{\rho}}$. Also important to state clearly is that, all 
the above implies that the gravitational field is what influences the distribution of matter -- this, in our view, resonates both with logic and 
intuition. We shall demonstrated the assertion that: 
$\rho(r)=\int^{r}_{0}\int^{2\pi}_{0}r^{2}\rho(r,\theta)\sin\theta d\theta dr\propto r^{-\alpha_{\rho}}$. We know that:

\begin{equation}
\int^{r}_{0}\int^{2\pi}_{0}r^{2}\rho(r,\theta)\sin\theta d\theta dr=\int^{r}_{0}r^{2}\left(\int^{2\pi}_{0}\rho(r,\theta)\sin\theta d\theta\right) dr=
4\pi\int^{r}_{0}r^{2}\rho(r) dr
\end{equation}

\noindent this means:

\begin{equation}
\rho(r)=\frac{1}{4\pi}\int^{2\pi}_{0}\rho(r,\theta)\sin\theta d\theta.
\end{equation}

\noindent Our claim is that if $\rho(r,\theta)$ is given by (\ref{rhochoice}) such that $\Phi(r,\theta)$ is given by (\ref{newpot2}), where 
$\mathcal{M}(r)$ in (\ref{newpot2}) is such that $\mathcal{M}(r)\propto r^{\alpha}$ for some constant $\alpha$, then:

\begin{equation}
\rho(r)=\frac{1}{4\pi}\int^{2\pi}_{0}\rho(r,\theta)\sin\theta d\theta\propto r^{\alpha_{\rho}}\label{claim},
\end{equation}

\noindent where $\alpha_{\rho}$ is some constant. We know that:

\begin{equation}
\frac{1}{4\pi}\int^{2\pi}_{0}\rho(r,\theta)\sin\theta d\theta=-\frac{1}{16\pi^{2} G}\int^{2\pi}_{0}\left(\frac{2}{r}\frac{\partial}{\partial r }-
\frac{1}{r^{2}}\right)\frac{\partial \Phi(r,\theta)}{\partial\theta}\sin\theta d\theta. 
\end{equation}

\noindent We have substituted $\rho(r,\theta)$ in (\ref{rhochoice}) into the above. This simplifies to:

\begin{equation}
\frac{1}{4\pi}\int^{2\pi}_{0}\rho(r,\theta)\sin\theta d\theta=-\frac{1}{16\pi^{2} G}\left(\frac{2}{r}\frac{\partial}{\partial r }-\frac{1}{r^{2}}\right)
\int^{2\pi}_{0}\frac{\partial \Phi(r,\theta)}{\partial\theta}\sin\theta d\theta. 
\end{equation}

\noindent From (\ref{newpot2}), we know that:

\begin{equation}
\frac{\partial \Phi(r,\theta)}{\partial\theta}=\sum^{\infty}_{\ell=0}\lambda_{\ell}c^{2}\left(\frac{G\mathcal{M}(r)}{rc^{2}}\right)^{\ell+1}
\sin\theta\frac{\partial P_{\ell}(\cos\theta)}{\partial (\cos\theta)}
\end{equation}

\noindent and this implies:

\begin{equation}
\frac{1}{4\pi}\int^{2\pi}_{0}\rho(r,\theta)\sin\theta d\theta =-\frac{c^{2}}{16\pi^{2} G}\left(\frac{2}{r}\frac{\partial}{\partial r }-
\frac{1}{r^{2}}\right)\sum^{\infty}_{\ell=0}\lambda_{\ell}\int^{2\pi}_{0}\left(\frac{G\mathcal{M}(r)}{rc^{2}}\right)^{\ell+1}\sin^{2}\theta
\frac{\partial P_{\ell}(\cos\theta)}{\partial (\cos\theta)}d\theta
\end{equation}

\noindent this simplifies to:

\begin{equation}
\frac{1}{4\pi}\int^{2\pi}_{0}\rho(r,\theta)\sin\theta d\theta =-\frac{c^{2}}{16\pi^{2} G}\left(\frac{2}{r}\frac{\partial}{\partial r }-
\frac{1}{r^{2}}\right)\sum^{\infty}_{\ell=0}\lambda_{\ell}\left(\frac{G\mathcal{M}(r)}{rc^{2}}\right)^{\ell+1}\overbrace{\int^{2\pi}_{0}\sin^{2}\theta  \frac{dP_{\ell}(\cos\theta)}{d(\cos\theta)}d\theta}^{\textrm{Let}\,\, \textrm{this}\,\, \textrm{be:}\,\, \textrm{I}_{\ell}\textrm{(}\theta\textrm{)}}
\end{equation}

\noindent where $I_{\ell}(\theta)$ is as defined above. It should not be difficult to see that  $I_{0}(\theta)=0$, $I_{1}(\theta)=1$ and that 
$I_{\ell}(\theta)\equiv0$ for all $\ell\geq2$. From this, it follows that:

\begin{equation}
\rho(r)=\frac{1}{4\pi}\int^{2\pi}_{0}\rho(r,\theta)\sin\theta d\theta = \left(\frac{\lambda_{1}c^{2}}{16\pi^{2} G}\right)\left(\frac{2}{r}
\frac{\partial}{\partial r }-\frac{1}{r^{2}}\right)\left(\frac{G\mathcal{M}(r)}{rc^{2}}\right)^{2},
\end{equation}

\noindent Now, if $\mathcal{M}(r)\propto r^{\alpha}$ this means $\mathcal{M}(r)=k r^{\alpha}$ for some adjustable constant $k$. Plugging this into the 
above, one obtains:

\begin{equation}
\rho(r)= \left(\frac{\left(2\alpha-1\right)\lambda_{1}c^{2}}{16\pi^{2} G}\right)\left(\frac{Gk}{c^{2}}\right)^{2}r^{2\alpha-4}.
\end{equation}

\noindent This\footnote{Under the prescribed conditions $\mathcal{M}(r)\propto r^{\alpha}$ leads to $\rho(r)\propto r^{2\alpha-4}$.  While 
$\mathcal{M}(r)=\int^{r}_{0}\int^{2\pi}_{0}\rho(r,\theta)\sin\theta d\theta dr$,  the basic definition $\mathcal{M}(r)=4\pi r^{3}\rho(r)/3$ must hold 
too, since $\mathcal{M}(r)$ is the amount of mass enclosed in volume sphere of radius $r$ and $\rho(r)$, is the mass-density of material in this volume 
sphere. These two definitions must  lead to identical formulas. If this is to be so -- then;  one is lead to the conclusion that $\alpha=1$, and this 
means $\mathcal{M}(r)\propto r$ and $\rho(r)\propto r^{-2}$. In the face of observations, the later result is very interesting since MCs seem to favor 
this density profile.} verifies our claim in (\ref{claim}). As already said, all the above implies that the gravitational field is what influences the 
distribution of matter. Co-joining this result with the result $(0\leq\alpha_{\rho}<3)$ in Nyambuya (\cite{nyambuya10c}) (hereafter Paper III), it 
follows that $(0.5\leq\alpha<2)$. Further,  a deduction to be made from the above result is that the spin does control the mass distribution 
\textit{via} the term $\lambda_{1}$.

\section[\textbf{The Undetermined Constants $\lambda_{\ell}$}]{\label{undconst} The Undetermined Constants $\lambda_{\ell}$}

Again, as already stated in Paper I, one of the draw backs of the ASTG is that it is heavily dependent on observations \textit{for} the values of 
$\lambda_{\ell}$ have to be determined from observations. Without knowledge of the $\lambda_{\ell}'s$, one is unable to produce the hard numbers 
required to make any numerical quantifications. Clearly, a theory incapable of making any numerical quantifications is -- in the physical realm, 
useless. To avert this, already in Paper I and as-well in Nyambuya (\cite{nyambuya10b}) (hereafter Paper II) an effort to solve this problem has been 
made. In Paper I, a  \textit{reasonable suggestion} was made to the effect  that:

\begin{equation}
\lambda_{\ell}=\left(\frac{(-1)^{\ell+1}}{\left(\ell^{\ell}\right)!\left(\ell^{\ell}\right)}\right)\lambda_{1}.
\label{oldlambda}
\end{equation}

\noindent This suggestion meets the intuitive requirements stated there-in Paper I.  If these $\lambda$'s are to be given by (\ref{oldlambda}), then, 
there is just one unknown parameter and this parameter is $\lambda_{1}$. The question is what does this depend on? We strongly feel/believe that 
$\lambda_{1}$ is dependent on the spin angular frequency and the radius of the gravitating body in question and our reasons are as follows.

The ASTG will be shown shortly to be able to explain outflows as a gravitational phenomenon. Pertaining to their association with star formation 
activity, it is believed that molecular outflows are a necessary part of the star formation process because their existence may explain the apparent 
angular momentum imbalance. It is well known that the amount of initial angular momentum in a typical star-forming cloud core is several orders of 
magnitude too large to account for the observed angular momentum found in formed or forming stars (see \textit{e.g.} Larson \cite{larson03b}). The 
sacrosanct Law \textit{of} Conservation of angular momentum informs us that this angular momentum can not just disappear into the oblivion of 
interstellar spacetime. So, the question is where does this angular momentum go to? It is here that outflows are thought to come to the rescue as they 
can act as a possible agent that carries away the excess angular momentum. Whether or not this assertion is true or may have a bearing with reality, no 
one really knows.

This angular momentum, if it where to remain as part of the nascent star, it would, \textit{via} the  strong centrifugal forces (the centrifugal 
acceleration is given by: $a_{c}=\omega^{2}_{star}\mathcal{R}_{star}$),  tear the star apart.  This however does not explain, why they [outflows]  
exist and how they come to exist but simple posits them as a vehicle needed to explain the mystery of ``{The Missing Angular Momentum Problem}'' in 
star forming systems and the existence of stars in their intact and compact form as firery balls of gas.

In Paper II, guided more by intuition than anything else, it was drawn  from the tacit thesis ``that outflows possibly save the star from the 
detrimental centrifugal forces'',  the suggestion that $\lambda_{1}\propto (a_{c})^{\zeta_{0}}$ where $\zeta_{0}$ is a pure constant that must be 
universal, that is, it must be the same for all spinning gravitating systems. This suggestion, if correct leads us to:

\begin{equation}
\lambda_{\ell}=\left(\frac{(-1)^{\ell+1}}{\left(\ell^{\ell}\right)!\left(\ell^{\ell}\right)}\right)\left(\frac{a_{c}}{a_{*}}\right)^{\zeta_{0}}
\label{newlambda1}.
\end{equation}

\noindent Knowing the solar values of $\lambda_{1}$ and as-well the value of $\zeta_{0}$, one is lead to: $a_{*}=\omega^{2}_{\odot}\mathcal{R}_{\odot}
(\lambda_{1}^{\odot})^{-\frac{1}{\zeta_{0}}}$. As will be demonstrated soon, the term $\lambda_{1}$ controls outflows. Given that $\lambda_{1}$ controls 
outflows and that outflows possibly aid the star in shedding off excess spin angular momentum, the best choice\footnote{We speak of ``choice'' here as 
though the decision is ours on what this parameter must be. No, the decision was long made by \textit{Nature}, ours is to find out what choice 
\textit{Nature} has made. That said,  we should say that, this ``choice'' is made with expediency -- \textit{i.e.}, this choice which is based on 
intuition, is to be measured against experience.} for this parameter is one that leads to these outflows responding to the spin of the star and as well 
the centrifugal forces generated by this spin in such a way that the star is able to shed off this excess spin angular momentum. So, what led to this 
proposal $\lambda_{1}\propto (a_{c})^{\zeta_{0}}$ is the aforesaid.  Now, we shall revise this suggestion by advancing what we believe is a far much 
better argument.

If outflows are there to save the nascent star from the ruthlessness of the centrifugal forces,  then, it is logical  to imagine that at the moment the 
centrifugal forces are about to rip the star apart, outflows will switch-on, thus shedding off this excess spin angular momentum. The centrifugal forces
 have their maximum toll on the equatorial surface of the star hence if the centrifugal forces are to rip the nascent star apart, this would start at 
the equator of the nascent star. The centrifugal force on the surface of the star acting on a particle of mass $m$ is 
$F_{c}=m\omega^{2}_{star}\mathcal{R}_{star}=ma_{c}$  and the gravitational force on the same particle is 
$F_{g}=G\mathcal{M}m/\mathcal{R}_{star}^{2}=mg_{star}$. Now lets define the quotient $\mathcal{Q}=F_{c}/F_{g}=a_{c}/g_{star}$. If the particle where to 
stay put on the surface of the star, then we will have $F_{c}-F_{g}<0\Rightarrow\mathcal{Q}<1$; and if the particle where to fly \textit{off} the 
surface, we will have $F_{c}-F_{g}>0\Rightarrow\mathcal{Q}>1$. The critical condition before the star begins to be torn apart is 
$F_{c}-F_{g}=0\Rightarrow\mathcal{Q}=1$. All the above can be summarized as:

\begin{equation}
\mathcal{Q}:=\left\{\begin{array}{l l l}<1 & & \textrm{No\,\, Outflow\,\,Activity}\\ =1 & &  \textrm{Critical\,\, Condition} \\ > 1 & & \
textrm{Outflow\,\, Activity}\end{array}\right. .
\end{equation}

\noindent Lets call this quotient, the Outflow Control Quotient (OCQ). Clearly, the OCQ determines the necessary conditions for outflows to switch on. 
Given this, and as-well the thinking that $\lambda_{1}$ controls outflows, the suggestion is clear that $\lambda_{1}\propto \mathcal{Q}^{\zeta_{0}}$. If 
this is correct, then:

\begin{equation}
\lambda_{1}=\zeta\mathcal{Q}^{\zeta_{0}}\label{lambda-rel}.
\end{equation}

\noindent We shall take this as our proposal for $\lambda_{1}$ and this means we must determine $(\zeta,\zeta_{0})$. From the above, it follows that:

\begin{equation}
\frac{\lambda_{1}^{\oplus}}{\lambda_{1}^{\odot}}=\left(\frac{\mathcal{Q}_{\oplus}}{\mathcal{Q}_{\odot}}\right)^{\zeta_{0}},
\end{equation}

\noindent where $\mathcal{Q}_{\oplus}=a_{c}^{\oplus}/g_{\oplus}$ and $a_{c}^{\oplus}$ is the centripetal acceleration generated by the Earth's spin at 
the equator and $g_{\oplus}$ is the gravitational field strength at the Earth equator. Likewise, $\mathcal{Q}_{\odot}=a_{c}^{\odot}/g_{\odot}$, is the 
solar outflow quotient where $a_{c}^{\odot}$ is the centripetal acceleration generated by the Sun's spin at the solar equator and $g_{\odot}$ is the 
gravitational field strength at the solar equator.  Given that: $(\omega_{\oplus}=7.27\times10^{-5}\,\textrm{Hz}$ and 
$\omega_{\odot}=2.04\times10^{-5}\,\textrm{Hz})$, $(\mathcal{R}_{\oplus}=6.40\times10^{6}\,\textrm{m}$ and 
$\mathcal{R}_{\odot}=6.96\times10^{8}\,\textrm{m})$ and $(g_{\oplus}=9.80\,\textrm{ms}^{-2}$ and $g_{\odot}=27.9g_{\oplus})$. From this data, it follows 
that: 

\begin{equation}
\frac{\mathcal{Q}_{\oplus}}{\mathcal{Q}_{\odot}}=169.
\end{equation}

\noindent Now, in Paper II, we did show that depending on how one interprets the flyby equation, one obtains two values of $\lambda_{1}^{\oplus}$, 
\textit{i.e.}  $\lambda_{1}^{\oplus}=(2.00\pm0.80)\times10^{3}$ and $\lambda_{1}^{\oplus}=(1.50\pm0.70)\times10^{4}$. If the spin of the Earth is 
significantly variable during the course of its orbit around the Sun, we will have  $\lambda_{1}^{\oplus}=(2.00\pm0.80)\times10^{3}$ and if the spin is 
not significantly variable, then, $\lambda_{1}^{\oplus}=(1.50\pm0.70)\times10^{4}$. If $\lambda_{1}^{\oplus}=(2.00\pm0.80)\times10^{3}$, then:

\begin{equation}
\frac{\lambda_{1}^{\oplus}}{\lambda_{1}^{\odot}}=\frac{15000\pm7000}{21.00\pm4.00}=800\pm500,\label{outfratio1}
\end{equation}

\noindent and from this it follows that $800\pm500=169.19^{\zeta_{0}}$, hence $\zeta_{0}=1.30\pm0.10$. If $\lambda_{1}^{\oplus}=(1.50\pm0.70)
\times10^{4}$, then:  

\begin{equation}
\frac{\lambda_{1}^{\oplus}}{\lambda_{1}^{\odot}}=\frac{2000\pm800}{21.00\pm4.00}=100\pm60,\label{outfratio2}
\end{equation}

\noindent and from this it follows that $100\pm60=169.19^{\zeta_{0}}$ hence $\zeta_{0}=0.90\pm0.10$.

If $\zeta_{\oplus}$ and $\zeta_{\odot}$ are the $\zeta$-values for the Earth and the Sun respectively, then, for $\lambda_{1}^{\oplus}=15000\pm7000$, we 
will have  $\zeta_{\oplus}=(3.40\pm2.70)\times 10^{7}$ and $\zeta_{\odot}=(8.00\pm4.00)\times 10^{10}$; and  for $\lambda_{1}^{\oplus}=2000\pm800$, we 
will have  $\zeta_{\oplus}=(3.40\pm2.70)\times 10^{7}$ and $\zeta_{\odot}=(8.00\pm4.00)\times 10^{10}$. Table \ref{table:zeta} is a self explanatory 
summary of all the above calculations.  The mean values of $\zeta$ for the case $\lambda_{1}^{\oplus}=(2.00\pm0.80)\times10^{3}$ is $\zeta=(8.00\pm1.00)
\times10^{5}$ and for the case $\lambda_{1}^{\oplus}=(1.50\pm0.70)\times10^{4}$ is $\zeta=(4.00\pm2.00)\times10^{7}$. These mean values have been 
obtained by taking the values of $\zeta_{\oplus}$ and $\zeta_{\odot}$  where they intersect in their error margins. 

\begin{table}

\centering
\caption[\textbf{The $(\zeta_{0},\zeta)$ Values for the Two Different Values of $\lambda_{1}^{\oplus}$}]{\textbf{: The $(\zeta_{0},\zeta)$ Values for 
the Two Different Values of $\lambda_{1}^{\oplus}$.}}
\begin{tabular}{l r r r r}
\hline\\
$\lambda_{1}^{\oplus}$ & $\lambda_{1}^{\odot}$ & $\zeta_{0}$ & $\zeta_{\odot}$ & $\zeta_{\oplus}$ \\
 ($10^{3}$) & & & ($10^{5}$) & ($10^{5}$) \\
 \hline\hline\\
$\,\,\,2.00\pm0.80$   & $21.00\pm4.00$ & $0.90\pm0.10$ &  $13.00\pm6.00$ & $5.00\pm3.00$\\
$15.00\pm7.00$ & $21.00\pm4.00$ & $1.30\pm0.10$ & $500\pm400$  & $400\pm200$\\
\hline
\end{tabular}

\label{table:zeta}
\end{table}

As argued in Paper II, the value $\lambda_{1}^{\oplus}=(2.00\pm0.80)\times10^{3}$ has been obtained from the assumption that the spin of the Earth 
varies widely during its course on its orbit around the Sun. This is not supported by observations thus we are not persuaded to take-up/recommend this 
value of $\lambda_{1}^{\oplus}=(2.00\pm0.80)\times10^{3}$.  Also, as argued in Paper II, the value $\lambda_{1}^{\oplus}=(1.50\pm0.70)\times10^{4}$ is 
obtained from the assumption that the spin of the Earth does not vary widely during its course on its orbit. Thus, we shall adopt the values of 
$(\zeta_{0},\zeta)$ that conform with $\lambda_{1}^{\oplus}=(1.50\pm0.70)\times10^{4}$ and $\lambda_{1}^{\odot}=21.0\pm0.40$, hence:

\begin{equation}
\lambda_{1}=(4.00\pm2.00)\times10^{7}\left(\frac{a_{c}}{g_{star}}\right)^{1.30\pm0.10}\label{eqn:adoptedlambda1}.
\end{equation}

\noindent Obviously, the greatest criticism against this result is that it is obtained from just two data points. To obtain something more reliable, one 
needs more data points. This is something that a future study must handle, at present, we simple want to set-up the mathematical model from the little 
available data and when data becomes available, amendments are made accordingly. While we have used the minimal possible data points, one thing that can 
be deduced from this data is that this result obtained points to a correlation as proposed in (\ref{lambda-rel}) -- otherwise, if there was no 
correlation as proposed, the values of $(\zeta_{0},\zeta)$ obtained the two values of $\lambda_{1}$ do not vary widely as is expected if the proposed 
relationship (\ref{lambda-rel}) did not hold at all.

\section{Outflows as a Gravitational Phenomenon}

We shall look into the empty and non-empty space solution of the solution of the Poisson-Laplace equation and show that both these solutions exhibit a 
repulsive bipolar gravitational field and that this repulsive gravitational field is controlled by the parameter $\lambda_{1}$.

\subsection{Non-Empty Space Solutions}

Now, if one accepts what has been presented thus far -- as will be shown in this section; it follows that outflows may-well be a gravitational 
phenomena. First, from the previous section, it follows that we must take the ASTG only up to second order, \textit{i.e.}:

\begin{equation}
\Phi=-\frac{G\mathcal{M}(r)}{r}\left[1+\frac{\lambda_{1}G\mathcal{M}(r)\cos\theta}{rc^{2}}+\lambda_{2}\left(\frac{G\mathcal{M}(r)}{rc^{2}}\right)^{2}
\left(\frac{3\cos^{2}\theta-1}{2}\right)\right].\label{pot3}
\end{equation}

\noindent  We know that the gravitational field intensity: $\vec{\textbf{g}}(r,\theta)=-\boldsymbol{\nabla}\Phi(r,\theta)=
g_{r}(r,\theta)\hat{\textbf{r}}+g_{\theta}(r,\theta)\hat{\boldsymbol{\theta}}$, this means:

\begin{equation}
g_{r}=g_{N}\left[\overbrace{1+\frac{2\lambda_{1}G\mathcal{M}(r)\cos\theta}{rc^{2}}}^{\textbf{Term I}}+
\overbrace{3\lambda_{2}\left(\frac{G\mathcal{M}(r)}{rc^{2}}\right)^{2}\left(\frac{3\cos^{2}\theta-1}{2}\right)}^{\textbf{Term II}}\right],
\label{rgfield}
\end{equation}

\noindent where: $g_{N}=-G\mathcal{M}(r)/r^{2}$, is the Newtonian gravitational field intensity and:

\begin{equation}
g_{\theta}=g_{N}r^{2}\sin\theta\left[\frac{\lambda_{1}G\mathcal{M}(r)}{rc^{2}}+9\lambda_{2}\left(\frac{G\mathcal{M}(r)}{rc^{2}}\right)^{2}
\cos\theta\right]\label{tgfield}.
\end{equation}  

\noindent For gravitation to be exclusively attractive (as is expected), we must have:  [$g_{r}(r,\theta)>0$] and [$g_{\theta}(r,\theta)~>~0$]. From 
(\ref{rgfield}) and (\ref{tgfield}), it is clear that regions of exclusively repulsive gravitation will exist  and these will occur in the region where: 
[$g_{r}(r,\theta)<0$] and [$g_{\theta}(r,\theta)~<~0$]. This region where gravity is exclusively repulsive is the region where it is not attractive, it 
is the negated region of the region of attractive gravitation: [\textit{i.e.} $\left\{g_{r}(r,\theta)>0\right\}$ and 
$\left\{g_{\theta}(r,\theta)~>~0\right\}$].  Let us start by treating the case: [$g_{r}(r,\theta)~<~0$]. From (\ref{rgfield}), if: 
[$g_{r}(r,\theta)<0$], then: ({Term I} $<0$) and ({Term II} $<0$), as well. The condition: ({Term I} $<0$),  implies:

\begin{equation}
r<-\lambda_{1}\left(\frac{2G\mathcal{M}(r)}{c^{2}}\right)\cos\theta=\lambda_{1}\left(\frac{2G\mathcal{M}(r)}{c^{2}}\right)\cos\theta\label{lob},
\end{equation}

\noindent (NB: $\cos\theta\equiv-\cos\theta$) and if one where to take $r$ such that it only takes positive values, then, (\ref{lob}) must be written in 
the equivalent form:

\begin{equation}
r<\lambda_{1}\left(\frac{2G\mathcal{M}(r)}{c^{2}}\right)\left|\cos\theta\right|\label{alob},
\end{equation}

\noindent where the brackets $\left|[]\right|$ represents the absolute value. We have to explain this, \textit{i.e.} why we concealed the negative sign 
in (\ref{lob}) and inserted the absolute value operator in (\ref{alob}).  From (\ref{lob}), it is seen that this inequality includes negative values of 
$r$ and to avoid any confusion as to what these negative values of $r$ really mean, this needs to be explained for failure to do so or failure by the 
reader to understand this means they certainly will be unable to agree with the outflow ``picture'' laid down herein. This explanation is important in 
order to understand the morphology of the outflow and as-well the ASGF. 

For a moment, imagine a flat Euclidean plane and on this plane let $\textrm{O}$, $\textrm{A}$ and $\textrm{P}$ be distinct and separate points on this 
plane with $\textrm{O}$ and $\textrm{A}$ being fixed and $\textrm{P}$ is a variable point. In polar coordinates, as in the present case, a point 
$\textrm{P}$ is characterized by two numbers: the distance $(r\geq{0})$ to the fixed pole or origin $\textrm{O}$, and the angle $\theta$ the line 
$\textrm{OP}$ makes with the fixed reference line $\textrm{OA}$. The angle $\theta$  is only defined up to a multiple of ${360}\degree$ 
(or  ${2}\pi\,\textrm{rad}$, in radians). This is the conventional definition. Sometimes it is convenient as in the present case to relax the condition 
$(r\geq0)$ and allow $r$ to be assigned a negative value such that the point $(r,\theta)$ and $(-r, \theta +{180}\degree)$ represent the same-point, 
hence thus when ever we have $(-r,\theta)$  this must be replaced by $(r, \theta -{180}\degree)$. It is easier for us to always think of $r$ as always 
being positive. To achieve this, given the fact that $(-r,\theta)\equiv (r, \theta -{180}\degree)$, we must write (\ref{lob}) as has been done in 
(\ref{alob}),  hence (\ref{alob}) finds justification. This explanation can be found in any good mathematics textbook that deals extensively with polar 
coordinates. Hereafter, whenever a similar scenario arises where negative values of $r$ emerge, we will automatically and without notification assume 
that $(-r,\theta)$  is $(r, \theta -{180}\degree)$ and this will come with the introduction of the absolute value sign as has been done in (\ref{alob}).

Now, proceeding from where we left. As has already been explained at the beginning of this section, we have to substitute the Mass Distribution Function 
(MDF)  $\mathcal{M}(r)$  into (\ref{alob}) and having done so we would have to make $r$ the subject. It has been argued in  equation $24$ of Paper III, that for a MC that exhibits a density profile:  $\rho(r)\propto r^{-\alpha_{\rho}}$, where $\alpha_{\rho}$ 
is the density index, that the MDF is given by: 

\begin{equation}
\mathcal{M}(r)= \overbrace{\mathcal{M}_{csl} \left(\frac{r^{3-\alpha_{\rho}}-\mathcal{R}_{star}^{3-\alpha_{\rho}}}{\mathcal{R}^{3-\alpha_{\rho}}_{core}-
\mathcal{R}_{star}^{3-\alpha_{\rho}}}\right)}^{\textrm{\tiny \bf {Circumstellar}\,{Mass}\,{Inside}\,{Region}\,{of \,\,Radius}\,{r}}}+
\overbrace{\mathcal{M}_{star}}^{\textbf{ \textrm{\tiny Nascent\,Star's\,Mass}}}\,\,\,\,\,\,\,\,\,\textrm{for}\,\,\,\,\,\,r\geq\mathcal{R}_{star},
\end{equation}

\noindent where $\mathcal{M}_{csl}$ is total mass of the circumstellar material at any given time, $\mathcal{R}_{star}$ is the radius of the nascent star
 at any given time, $\mathcal{R}_{core}$ is the radius at any given time of the gravitationally bound core from which the star is forming.

Now,  substituting  the MDF (given above)  into (\ref{alob}) and thereafter making $r$ the subject of the formula  would lead to a horribly complicated 
inequality that would require the use of the Newton-Ralphson approach to solve. Since ours in the present is but a qualitative analysis, we can make 
some very realistic simplifying assumptions that can make our life much easier. If the spatial extent of the star is small compared to that of the core 
\textit{i.e.}: $(\mathcal{R}_{star}\ll\mathcal{R}_{core}\Rightarrow \mathcal{R}_{core}^{3-\alpha_{\rho}}-\mathcal{R}_{star}^{3-\alpha_{\rho}}
\simeq\mathcal{R}_{core}^{3-\alpha_{\rho}})$ and the mass of the star is small compared to the mass of the core \textit{i.e.}: 
$(\mathcal{M}_{star}\ll\mathcal{M}_{core}\Rightarrow \mathcal{M}_{csl}\simeq\mathcal{M}_{core})$, then, the MDF simplifies to:

\begin{equation}
\mathcal{M}(r)\simeq\mathcal{M}_{core}\left(\frac{r}{\mathcal{R}_{core}}\right)^{3-\alpha_{\rho}}\label{mass1}. 
\end{equation}

\noindent Inserting this into (\ref{alob}) and thereafter performing some basic algebraic computations that see $r$ as the subject of the formula, one 
is lead to:

\begin{equation}
r<\left[\lambda_{1}\left(\frac{2G\mathcal{M}_{core}}{c^{2}\mathcal{R}_{core}}\right)\mathcal{R}_{core}\right]\left|\cos\theta\right|^{\frac{1}{2-
\alpha_{\rho}}}\label{alob1}.
\end{equation}

\noindent Now, if we set:

\begin{equation}
\epsilon_{1}^{core}=\left(\left[\lambda_{1}\left(\frac{2G\mathcal{M}_{core}}{c^{2}\mathcal{R}_{core}}\right)\right]^{\frac{1}{2-\alpha_{\rho}}}\right)
\left(\frac{\mathcal{R}_{core}}{\mathcal{R}_{star}}\right),\label{epsiloncore}
\end{equation}

\noindent then (\ref{alob1}) reduces to:

\begin{equation}
r<\epsilon_{1}^{core}\mathcal{R}_{star}\left|\cos\theta\right|^{\frac{1}{2-\alpha_{\rho}}}=l_{max}\left|\cos\theta\right|^{\frac{1}{2-\alpha_{\rho}}}
\label{curve},
\end{equation}

\noindent where: $l_{max}=\epsilon_{1}^{core}\mathcal{R}_{star}$. On the \textit{xy}-plane as shown in figure \ref{outfdiag}, the equation: 
$r=l_{max}\left|\cos\theta\right|^{\frac{1}{2-\alpha_{\rho}}}$, describes two lobs. For the purposes of this reading, let the volume of revolution of 
the lob be called a loboid, and the loboid above the \textit{x}-axis shall be called the upper loboid, and likewise the loboid below the \textit{x}-axis 
shall be called the lower loboid.

Now,  the condition: ({Term II} $<0$), implies: $[\theta~<~\cos^{-1}(\pm1/\sqrt{3})]$, which means: $(-54.7<\theta<54.7)$. Now, for the azimuthal 
component to be repulsive, we must have: $[g_{\theta}(r,\theta)>0]$, we will have from (\ref{tgfield}), the condition:

\begin{equation}
r>-\left(\frac{9\lambda_{2}}{2\lambda_{1}^{2}}\right)\left(\frac{2\lambda_{1}G\mathcal{M}(r)}{c^{2}}\right)\cos\theta.\label{lobmin}
\end{equation}

\noindent Now going through the same procedure as above, (\ref{lobmin}) can be written as:

\begin{equation}
r> l_{min}\left|\cos\theta\right|^{\frac{1}{2-\alpha_{\rho}}},\label{lobmin1}
\end{equation}

\noindent where:

\begin{equation}
l_{min}=\left(\left|\frac{9\lambda_{2}}{2\lambda_{1}^{2}}\right|^{\frac{1}{2-\alpha_{\rho}}}\right)l_{max}\label{lmax}.
\end{equation}

\noindent Thus, coalescing the results, invariably, one is led to conclude that the region of repulsive gravitation is:  

\begin{equation}
\left[l_{min}< r< l_{max}\right]\,\textit{\&}\,\left[\cos^{-1}\left(-\frac{1}{\sqrt{3}}\right)<\theta<\cos^{-1}\left(\frac{1}{\sqrt{3}}\right)\right].
\label{repul}
\end{equation}

\noindent In the region described above, the gravitational field is both radially and azimuthally repulsive, that is, there is complete gravitational 
repulsion in this region. Pictorially, a summary of the  emergent picture of the repulsive gravitational field in shown in figure \ref{outfdiag}. This 
picture -- in our view, fits the description of outflows, the limiting factors are the sizes of $l_{max}$ and $l_{min}$, these values all depend on  the 
one parameter $\lambda_{1}$, hence thus, this parameter is the crucial parameter  which determines the properties of outflows. Shortly, we will discuss 
this picture but before this, it is necessary that we go through the empty space solutions as-well.

\begin{figure}
\centering
\shadowbox{\includegraphics[scale=0.6]{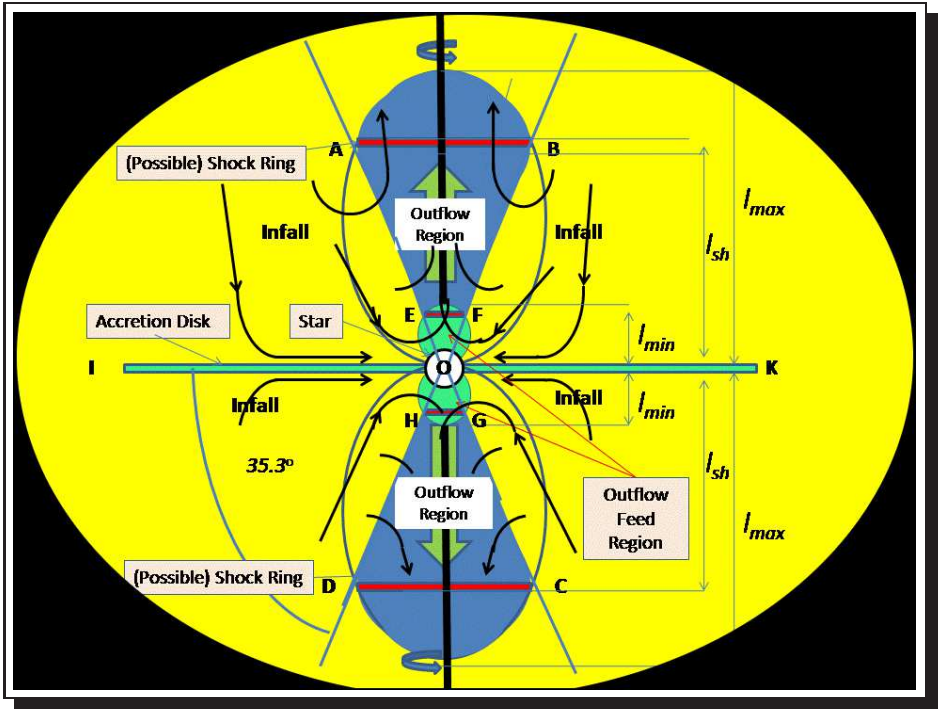}}
\caption[\textbf{Outflow Diagram}]{This figure illustrates the emergent picture from the azimuthally symmetric considerations of the Poisson equation. 
While fanning out matter in the region of repulsive gravitation, the rotating star is surrounded by an equatorial disk; once the outflow switches-on, 
this disk is the only channel \textit{via} which the mass of the star feeds. The disk is not affected by radiation in the sense that some of its 
material close to the nascent star will be swept away by the radiation field, no! The force of gravity along this disk is purely radial and is directed 
toward the nascent.}
\label{outfdiag}
\end{figure}

\subsection{Empty Space Solutions}

As will be demonstrated in this section, the picture imaging from the empty space solution is not different from that of the non-empty space solution. 
However, there is an important difference between these two pictures and this difference need to be stated. If our spinning gravitating body is  not 
giving off material like the Sun, then the  region of repulsive gravitation will occur inside the this body. We shall consider the star to be a point 
mass \textit{i.e.}, all of its mass is concentrated at the star's center of mass.

As before, from (\ref{rgfield}) and (\ref{tgfield}), it is clear that regions of repulsive gravitation will exist  and these will occur where 
[$g_{r}(r,\theta)<0$] and or [$g_{\theta}(r,\theta)~<~0$]. We shall as before start by treating the case [$g_{r}(r,\theta)~<~0$]. From (\ref{rgfield}), 
if [$g_{r}(r,\theta)<0$], then ({Term I} $<0$) and ({Term II} $<0$) as well. The condition ({Term I} $<0$)  implies:

\begin{equation}
r<-\lambda_{1}\left(\frac{2G\mathcal{M}}{c^{2}}\right)\cos\theta\label{e-lob},
\end{equation}

\noindent where in the present case $\mathcal{M}(r)$ must be replaced by $\mathcal{M}_{star}$ and this can be written in the equivalent form:

\begin{equation}
r<\lambda_{1}\left(\frac{2G\mathcal{M}_{star}}{c^{2}}\right)\left|\cos\theta\right|\label{e-alob}.
\end{equation}

\noindent Now, if we set:

\begin{equation}
\epsilon_{1}^{star}=\lambda_{1}\left(\frac{\mathcal{R}_{star}^{s}}{\mathcal{R}_{star}}\right),\label{e-epsilon}
\end{equation}

\noindent where $\mathcal{R}_{star}^{s}=2G\mathcal{M}_{star}/c^{2}$ is the Schwarzchild radius of the star, then (\ref{e-alob}) reduces to:

\begin{equation}
r<\epsilon_{1}^{star}\mathcal{R}_{star}|\cos\theta|=l_{max}|\cos\theta|\label{e-curve}.
\end{equation}

\noindent Now, the condition ({Term II} $<0$), as before, implies: $[\theta~<~\cos^{-1}(\pm1/\sqrt{3})]$ , which means: $(-54.7<\theta<54.7)$. Again as 
before, for the azimuthal component to be repulsive, we must have: $[g_{\theta}(r,\theta)>0]$, we will have from (\ref{tgfield}), that:

\begin{equation}
r>\left(\frac{9\lambda_{2}}{2\lambda_{1}^{2}}\right)\left(\frac{2\lambda_{1}G\mathcal{M}_{star}}{c^{2}}\right)\left|\cos\theta\right|,\label{-elobmin}
\end{equation}

\noindent and we need not explain anymore why the above can be written as:

\begin{equation}
r> l_{min}|\cos\theta|,\label{e-lobmin1}
\end{equation}

\noindent where this time:

\begin{equation}
l_{min}=\left|\frac{9\lambda_{2}}{2\lambda_{1}^{2}}\right|l_{max}\label{e-lmax}.
\end{equation}

\noindent Coalescing the results, invariably, one is led to conclude that the region of repulsive gravitation is:  

\begin{equation}
\left[l_{min}< r< l_{max}\right]\,\textit{\&}\,\left[\cos^{-1}\left(-\frac{1}{\sqrt{3}}\right)<\theta<\cos^{-1}\left(\frac{1}{\sqrt{3}}\right)\right].
\end{equation}

\noindent As in the case of the non-empty space, in the region described above, the gravitational field is both radially and azimuthally repulsive, 
hence there is complete gravitational repulsion in this region. The emergent picture is no different from that of the case of non-empty space. The 
important difference is that the region of gravitational repulsion is confined in the interior of the star if $(\epsilon_{1}<1)$, it is not visible 
outside. If $(\epsilon_{1}<1)$, there will exist no repulsive bipolar gravitational field that is  visible to beyond the surface of the spinning star. 
In the interior of the star, the solutions obtained for the case of non-empty space is what must apply.

\section[\textbf{ASGF of a Spinning Core with a Embedded Spinning  Star}]{ASGF of a Spinning Core with an Embedded Spinning  Star}

Central to the ASTG is that the material under consideration possesses some finite spin angular momentum. In the case of a nascent star embedded inside 
a gravitationally bound core, we are going to have the star's spin angular frequency being different to that of the circumstellar material; because, in 
the early stages when the nascent star is forming, the spin angular frequency of the circumstellar material and the star will, on the average, be the 
same since it is expected that circumstellar material and the star will co-rotate; but, because of the increasing mass and spin angular momentum of the 
nascent star due to the accretion of material, at some-point, the star must break-off from this co-rotational motion and spin independently of the 
circumstellar material, thus in the end, the star will have a different spin angular frequency to that of the circumstellar material. The different 
spin angular momentum of the nascent star and the circumstellar material will come along with  different $\lambda$-values. Assuming the circumstellar 
material is co-rotating with itself, it must have its own $\lambda$-value, let us call this $\lambda_{\ell}^{csl}$ and that for the star be 
$\lambda_{\ell}^{star}$.

If there is a way of calculating the ASGF of the star at point $(r,\theta)$ and that of the circumstellar material at that same point $(r,\theta)$, then 
one will be able to calculate the resultant ASGF at any point $(r,\theta)$ because the gravitational field is here a scalar. Let $\Phi_{star}$ be the 
Azimuthally Symmetric Gravitational Potential (ASGP) of the star and that of the circumstellar material be $\Phi_{csl}$. Knowing $\Phi_{star}$ and 
$\Phi_{csl}$, clearly the resultant ASGP $\Phi_{eff}$ at any point $(r,\theta)$ is $\Phi_{eff}=\Phi_{star}+\Phi_{csl}$, hence one will be able to obtain 
the resultant ASGF. The ASGF of the star is not difficult to obtain, we already know that it must be given by:

\begin{equation}
\Phi_{star}=-\frac{G\mathcal{M}_{star}}{r}\left[1+\frac{\lambda_{1}^{star}G\mathcal{M}_{star}\cos\theta}{rc^{2}}+
\lambda_{2}^{star}\left(\frac{G\mathcal{M}_{star}}{rc^{2}}\right)^{2}\left(\frac{3\cos^{2}\theta-1}{2}\right)\right].\label{potstar}
\end{equation}

\noindent Now, we have to obtain the ASGF of a spinning core. The gravitational potential (\ref{pot3}) is the potential of star that is co-rotating 
with the circumstellar material. If we remove the central star from this gravitational potential what remains is the gravitational potential of a 
spinning core. Removing the central star from this potential means set $\mathcal{M}_{star}=0$, hence, the gravitational potential of a spinning core 
must be:

\begin{equation}
\Phi_{csl}=-\frac{G\mathcal{M}_{csl}(r)}{r}\left[1+\frac{\lambda_{1}^{csl}G\mathcal{M}_{csl}(r)\cos\theta}{rc^{2}}+
\lambda_{2}^{csl}\left(\frac{G\mathcal{M}_{csl}(r)}{rc^{2}}\right)^{2}\frac{3\cos^{2}\theta-1}{2}\right],\label{potcsl}
\end{equation}

\noindent where $\lambda_{\ell}^{csl}$ is the $\lambda_{\ell}$-value for the spinning circumstellar material and:

\begin{equation}
\mathcal{M}_{csl}(r)= \mathcal{M}_{csl} \left(\frac{r^{3-\alpha_{\rho}}-\mathcal{R}_{cav}^{3-\alpha_{\rho}}(t)}{\mathcal{R}^{3-\alpha_{\rho}}_{core}(t)-
\mathcal{R}_{cav}^{3-\alpha_{\rho}}(t)}\right)\,\,\textrm{for}\,\,r\geq\mathcal{R}_{cav}(t)\label{masscsl},
\end{equation}

\noindent is the circumstellar material enclosed in radius $r$. Now, as argued already: $\Phi_{eff}=\Phi_{star}+\Phi_{csl}$, thus adding these two 
potentials (\textit{i.e.} \ref{potstar} \textit{\&} \ref{potcsl}),  one obtains:

\begin{equation}
\Phi_{eff}(r,\theta)=-\sum^{\infty}_{\ell=0}c^{2}\left(\frac{G\left\{\lambda_{\ell}^{star}\mathcal{M}_{star}^{\ell+1}+\lambda_{\ell}^{csl}
\mathcal{M}_{csl}^{\ell+1}(r)\right\}^{\frac{1}{^{\ell+1}}}}{rc^{2}}\right)^{\ell+1}P_{\ell}(\cos\theta).\label{effpot1}
\end{equation}

\noindent This is the ASGP of a star that spins independently from its core. For convenience, we can write: $\mathcal{M}^{eff}_{\ell}(r)=
\left\{\lambda_{\ell}^{star}\mathcal{M}_{star}^{\ell+1}+\lambda_{\ell}^{csl}\mathcal{M}_{csl}^{\ell+1}(r)\right\}^{\frac{1}{^{\ell+1}}}$, and call this 
the effective gravitational mass for the $\ell^{th}$ gravitational-pole. By $\ell^{th}$ gravitational-pole, it shall be understood to mean the 
$\ell^{th}$-term in the gravitational potential term. This means the above can be written in the clearer and simpler form:

\begin{equation}
\Phi_{eff}(r,\theta)=-\sum^{\infty}_{\ell=0}c^{2}\left(\frac{G\mathcal{M}^{eff}_{\ell}(r)}{rc^{2}}\right)^{\ell+1}P_{\ell}(\cos\theta).\label{effpot2}
\end{equation}

\noindent To second order approximation, this potential is given by:

\begin{equation}
\Phi_{eff}=-\left(\frac{G\mathcal{M}_{0}^{eff}(r)}{r}\right)\left[1+\gamma_{1}\lambda_{1}^{star}\left(\frac{G\mathcal{M}_{1}^{eff}(r)
\cos\theta}{rc^{2}}\right)+\gamma_{2}\lambda_{2}^{star}\left(\frac{G\mathcal{M}_{2}^{eff}(r)}{rc^{2}}\right)^{2}\left(\frac{3\cos^{2}\theta-1}{2}\right)
\right],\label{pot-star-csl}
\end{equation}

\noindent where: $\gamma_{\ell}=\mathcal{M}_{\ell}^{eff}(r)/\mathcal{M}_{0}^{eff}(r)$. We shall assume this ASGP for a star that spins independently 
from its core.

\section[\textbf{Outflow Power}]{ Outflow Power}

Clearly, we do have from the ASTG regions of repulsive gravitation whose shape is similar to that seen in outflows. If these outflows are really powered 
by gravity, the question is: does the gravitational field have that much energy to drive these and if so, where does this energy come from? To answer 
this question, one will need to know the dominant radial component of the gravitational force since outflows dominantly operate along the radial 
direction. Clearly, one of the new extra poles in the gravitational field must be the cause of the outflows since without them, there are no outflow. 
For our investigations, the correct gravitational potential to use is (\ref{effpot2}) and of interest in this potential is the gravitational potential 
of the star. This invariably means we are looking at (\ref{potstar}).  So doing, one sees that the first order term (involving $\lambda_{1}$) is an 
all-repulsive term as already argued  while the second order them (involving $\lambda_{2}$) is repulsive and attractive, it depends on the region under 
consideration. 

Now, to ask what powers outflows amounts to asking: ``What is their energy source?''. If this energy source is the gravitational field, then, we know 
that the energy stored in the gravitational field whose potential is described by  $\Phi(r,\theta)$, is given by:

\begin{equation}
\mathcal{E}^{star}_{gpe}(r)=\int^{\mathcal{M}_{star}}_{0}\int^{\Phi(\infty,2\pi)}_{\Phi(r,0)}d\Phi(r,\theta)d\mathcal{M},
\end{equation}

\noindent and plugging into the above the ASGP, and thereafter performing the integration, one is led to:

\begin{equation}
\mathcal{E}^{star}_{gpe}(r)=-\frac{G\mathcal{M}_{star}^{2}}{2r}\left[1+\frac{\lambda_{1}G\mathcal{M}_{star}}{rc^{2}}+
\lambda_{2}\left(\frac{G\mathcal{M}_{star}}{rc^{2}}\right)^{2}\right],
\end{equation}

\noindent and using the fact that $\mathcal{L}_{star}=\mathcal{L}_{\odot}\left(\mathcal{M}_{star}/\mathcal{M}_{\odot}\right)^{3}$, one is further led 
to:

\begin{equation}
\mathcal{E}^{star}_{gpe}(r)=-\frac{G\mathcal{M}_{\odot}^{2}}{2r}\left(\frac{\mathcal{L}_{star}}{\mathcal{L}_{\odot}}\right)^{\frac{2}{3}}\left[1+
\frac{\lambda_{1}G\mathcal{M}_{\odot}}{rc^{2}}\left(\frac{\mathcal{L}_{star}}{\mathcal{L}_{\odot}}\right)^{\frac{1}{3}}\right].
\end{equation}

\noindent If $\mathcal{M}_{out}$ is the mass of the outflow at position $r$ and $V_{out}$ is the speed of this outflow at this position and 
$\mathcal{K}_{out}$ is the kinetic energy, we know that:

\begin{equation}
\left<\frac{d\mathcal{M}_{out}(r)}{dt}\right>=\frac{1}{V_{out}^{2}}\frac{d\left[\mathcal{M}_{out}(r)V_{out}^{2}\right]}{dt}=
\frac{2}{V_{out}^{2}}\frac{d\mathcal{K}_{out}}{dt},
\end{equation}

\noindent where the bracket $\left<[]\right>$ tells us that we are looking at the average. Now if the gravitational energy $\mathcal{E}^{star}_{gpe}(r)$ 
is equal to the kinetic energy of the outflow, then, from the above and, coupled with the said, one is led to:

\begin{equation}
\left<\frac{d\mathcal{M}_{out}(r)}{dt}\right>=-\frac{\tau_{G}G\mathcal{M}_{\odot}^{2}V_{out}^{-2}}{r}
\left(\frac{\mathcal{L}_{star}}{\mathcal{L}_{\odot}}\right)^{\frac{2}{3}}\left[1+\frac{2\lambda_{1}G\mathcal{M}_{\odot}}{rc^{2}}
\left(\frac{\mathcal{L}_{star}}{\mathcal{L}_{\odot}}\right)^{\frac{1}{3}}\right],\label{outfrel}
\end{equation}

\noindent where $\tau_{G}=\dot{G}/G$ and $\dot{G}$ is the time derivative of the Newton's gravitational constant.  In the derivation of the above, we 
have considered only first order terms and we have assumed that the gravitational constant is not a constant. Evidence that the gravitational constant 
maybe changing exists \textit{e.g.} see Pitjeva (\cite{pitjeva05}) and references therein. The ASTG also points to a variation of the gravitational constant and 
the details of this are  being worked out\footnote{We are at an advanced stage of preparation of this  work and it will soon be archived 
on \textit{viXra.org}: check Golden Gadzirayi Nyambuya's profile. Title of the Paper: \textit{A Foundational Basis for Variable-G and Variable-c 
Theories}.} and we give in the subsequent  paragraphs how this comes about.  

As it stands, the Poisson equation ($\vec{\nabla}^{2}\Phi=4\pi G\rho$) for a time varying $\Phi$ \textit{\&} $\rho$,  is not in conformity with the 
Relativity Principle. According to our current understanding of physics and \textit{Nature}, the seemingly sacrosanct Relativity Principle is a symmetry 
that every Law \textit{of} Physics must fulfill. The Relativity Principle states that Laws \textit{of} Physics must be independent of the observer's 
state of motion and as-well of the coordinate system used to formulate them. If the Poisson equation is to be a Law \textit{of} Nature, then, it must 
successfully fulfill the Relativity Principle. This means we must extend the Poisson equation to meet this requirement and the most natural and readily 
available such is:

\begin{equation}
\vec{\nabla}^{2}\Phi-\frac{1}{c^{2}}\frac{\partial^{2}\Phi}{\partial t^{2}}=4\pi G\rho,\label{4poisson}
\end{equation}

\noindent where $t$ is the time coordinate. This equation satisfies the Relativity Principle simply because it directly emerges from Einstein's equation 
of the General Theory \textit{of} Relativity (GTR). We know Einstein's GTR, specifically the Law \textit{of} Gravitation relating matter to the 
curvature of spacetime, does satisfy the  Relativity Principle; hence (\ref{4poisson}) too, satisfies the Relativity Principle. This equation 
(\textit{i.e.} \ref{4poisson}) is what we are working out, we shall show that it leads to a time variable $G$. So, as will be shown in the near future, 
the time variable $G$ in (\ref{outfrel}) is not without a basis.

Now, from (\ref{outfrel}), one sees that: $\dot{\mathcal{M}}_{out}(r)\propto V_{out}^{-2}(r)\mathcal{L}^{2/3}_{star}$. Given as stated in the 
introduction that observations find: $\dot{\mathcal{M}}_{out}(r)\propto V_{out}^{-1.8}\mathcal{L}^{0.6}_{star}$, which is close to what we have 
deduced here; this points to the fact that the thesis leading to our deduction: 
$\dot{\mathcal{M}}_{out}(r)\propto V_{out}^{-2}(r)\mathcal{L}^{2/3}_{star}$, may very well be  on  the right path of discovery. This clearly points to 
the need to look into these matters deeper than has been done here.

From the above, clearly -- a meticulous study of outflows should be able to measure the time variation in the gravitational constant $G$ and this hinges 
on the corrects of the ASTG. This would require higher resolution observations to measure the mass outflow rate [$\dot{\mathcal{M}}_{out}(r)$] at 
position $r$ from the star and as well the speed of the outflow at that point and knowing the mass or luminosity of the central driving source, a graph 
of, \textit{e.g.}: $r\dot{\mathcal{M}}_{out}(r)$ \textit{vs} $V_{out}^{-2}(r)\mathcal{L}^{2/3}_{star}$, should in accordance with the ideas above, 
produce a straight line graph whose slope is $\tau_{G}G\mathcal{M}_{\odot}^{2}$. This kind of work, if it where possible, it would help in making an 
independent confirmation of the measured time variation of Newton's constant of gravitation and it would act as further testing grounds for the 
falsification of the ASTG.

\section{Outflow Anatomy}

Briefly, we shall look into the anatomy of the outflow. We say ``briefly'' because each of the issues we shall look into requires a separate reading to 
fully address them. First, before we do that, it is important to find out when does the outflow switch-on and also when does it switch-off. That at 
some point in time in the evolution of a star, outflows switch-on and off is not debatable. So, before we even look into them, it makes perfect sense to 
investigate this. From figure \ref{outfdiag}, we see  that the anatomy of the outflow has been identified with four regions, \textit{i.e.} the 
\textit{Outflow Feed Region},  the \textit{Outflow Region} and the \textit{Shock Ring}. After  investigating the switching-on and off of the outflow, 
we will  look into the nature of these regions. Our analysis is qualitative rather than quantitative. We believe a quantitative analysis will require a 
fully-fledged numerical code. Work on this numerical code is underway.

\subsection{Switching-on \textit{of} Outflows}

Let us call the loboid described by (\ref{curve}) the outflow loboid and likewise the loboid described by (\ref{lobmin1}) the outflow feed loboid. From 
the preceding section, it is abundantly clear that we are going to have repulsive bipolar regions whose surface is described by a cone and a outflow 
loboid section. From this, we know that the maximum spacial extend of the repulsive gravitational field region will be given by the maximum spatial 
length of the lobes which occurs when: $\cos\theta=1$, \textit{i.e.} $l_{max}=\epsilon_{1}^{star}\mathcal{R}_{star}$. Now, to ask the question when 
does the outflow switch-on amounts to asking when is $l_{max}$ equal to the radius of the star? because the repulsive gravitational field will only 
manifest beyond the surface of the star \textit{if and only if} the maximum spatial extent of the region of repulsive gravitation is at least equal to 
the radius of the star, \textit{i.e.}: $l_{max}~\geq~\mathcal{R}_{star}$, this means, $l_{max}~=~\epsilon_{1}^{star}\mathcal{R}_{star}$; clearly, this 
will occur when: $(\epsilon_{1}^{star}=1)$. Therefore, outflows will switch-on when the condition: $(\epsilon_{1}=1)$, is reached, otherwise when: 
$(\epsilon_{1}^{star}<1)$, the repulsive gravitational field is confined inside the star.

This strongly suggests that if we are to use the ASGT to model outflows, then we must think of $\epsilon_{1}$ (hence $\lambda_{1}$) as an evolutionary
 parameter of the star \textit{i.e.}, this value starts of from a given absolute minimum value $($say $\epsilon_{1}^{star}=0)$, and as the star evolves, 
this value gets larger and larger until such a time that the repulsive gravitational field is switched on when: $(\epsilon_{1}^{star}=1)$, and 
thereafter it continues to grow and as it grows so does the spatial extend of the outflow (since this parameter controls the spatial size of the region 
of the repulsive gravitational field).

If the outflow switches on -- as it must, the  question is: ``Why does it switch on at that moment when it switches on and not at any other moment? 
What is so special about that moment when it  switches on that triggers it [outflows] to switch on?'' As we have already argued, this special moment is 
when $(\epsilon_{1}^{core}=1)$ for a star that co-rotates with its parent core and $(\epsilon_{1}^{star}=1)$ for a star that rotates independently of 
its parent core. From equation (\ref{epsiloncore} and \ref{e-epsilon}), this means we must have: 

\begin{equation}
\epsilon_{1}^{core}=\left[\zeta\left(\frac{4\pi^{2}\mathcal{R}_{core}^{3}}{G\mathcal{M}_{core}\mathcal{T}_{core}^{2}}\right)^{\zeta_{0}}
\left(\frac{2G\mathcal{M}_{core}}{c^{2}\mathcal{R}_{core}}\right)\right]^{\frac{1}{2-\alpha_{\rho}}}\left(\frac{\mathcal{R}_{core}}{\mathcal{R}_{star}}
\right)=1,
\end{equation}

\noindent and for a star that rotates independently of its core:

\begin{equation}
\epsilon_{1}^{star}=\zeta\left(\frac{4\pi^{2}\mathcal{R}_{star}^{3}}{G\mathcal{M}_{star}\mathcal{T}_{star}^{2}}\right)^{\zeta_{0}}
\left(\frac{2G\mathcal{M}_{star}}{c^{2}\mathcal{R}_{star}}\right)=1,
\end{equation}

\noindent where $(\mathcal{T}_{core}, \mathcal{T}_{star})$ are the period of the spin of the core and the star respectively. If 
$\mathcal{T}_{core}^{on}$ is the period of the core's spin when the outflow switches on and $\mathcal{T}_{star}^{on}$ is the period of the spin when 
the outflow switches on, then, from the above equations, it follows that: 

\begin{equation}
\mathcal{T}^{on}_{core}=\left(\frac{\pi}{c}\right)\left[\zeta\left(\frac{2G\mathcal{M}_{core}}{c^{2}}\right)^{1-\zeta_{0}}\left(\mathcal{R}_{star}^{on}
\right)^{\alpha_{\rho}-1}(\mathcal{R}_{core}^{on})^{3\zeta_{0}-\alpha_{\rho}+1}\right]^{\frac{1}{2\zeta_{0}}},
\end{equation}

\begin{equation}
\mathcal{T}^{on}_{star}=\left(\frac{\pi}{c}\right)\left[\zeta\left(\frac{2G\mathcal{M}_{star}^{on}}{c^{2}}\right)^{1-\zeta_{0}}
\left(\mathcal{R}_{star}^{on}\right)^{3\zeta_{0}-1}\right]^{\frac{1}{2\zeta_{0}}},
\end{equation}

\noindent where $(\mathcal{M}_{star}^{on},\mathcal{R}^{on}_{star}, \mathcal{R}^{on}_{core})$ are the mass and radius of the star and core at the time 
the outflow switches on respectively. From this, it follows that if the Sun were to spin on its axis once in every $7.70\pm0.40\,\textrm{hrs}$ 
(\textit{i.e.} $39.0\pm2.00\,\mu\textrm{Hz}$), the bipolar repulsive gravitational field must switch on and for the Earth, it would require it to spin 
once on its axis in every $10.00\pm2.00\,\textrm{min}$ (\textit{i.e.} $1.80\pm0.50\,\textrm{mHz}$). If the above is correct, then the Earth must spin 
about one hundred and forty four times its current spin  in order to achieve the bipolar repulsive gravitational field while the Sun must spin about 
five thousand six hundred its current spin rate to achieve a bipolar repulsive gravitation. The spin rate of the Earth is far less than that needed to 
cause the bipolar repulsive gravitational to switch on thus polar bears can smile knowing they will not fly off into space anytime soon.

We know that outflows are not always present, at some-point in the evolution of the star, they switch-off. What could cause them to do so? Given the 
reality that within the outflow loboid, there is the outflow feed loboid; this too, grows in size as the outflow loboid grows; at some-point the outflow 
and the outflow feed loboid will become equal -- leaving the outflow with no feed point. At this point when the outflow and outflow feed loboids become 
equal, clearly, the outflow must  switch-off. This occurs when $l_{max}=l_{min}$ and from (\ref{lmax}) this means the condition for this to occur is 
$|\lambda_{2}|=2\lambda_{1}^{2}/9$ and given that $\lambda_{2}=-\lambda_{1}/96$, this means $\lambda_{1}^{off}=9/192$. From (\ref{epsiloncore}), it 
follows that:

\begin{equation}
\lambda_{1}^{off}=\zeta\left(\frac{4\pi^{2}\mathcal{R}_{star}^{3}}{G\mathcal{M}_{star}\mathcal{T}_{star}^{2}}\right)^{\zeta_{0}}
\left(\frac{2G\mathcal{M}_{star}}{c^{2}\mathcal{R}_{star}}\right)=\frac{9}{192},
\end{equation}

\noindent this implies:

\begin{equation}
\mathcal{T}^{off}_{star}=\left(\frac{192}{9}\right)^{\frac{1}{2\zeta_{0}}}\left(\frac{\pi}{c}\right)\left[\zeta
\left(\frac{2G\mathcal{M}_{star}^{off}}{c^{2}}\right)^{1-\zeta_{0}}\left(\mathcal{R}_{star}^{off}\right)^{3\zeta_{0}-1}\right]^{\frac{1}{2\zeta_{0}}},
\end{equation}

\noindent where likewise $(\mathcal{M}_{star}^{off},\mathcal{R}^{off}_{star})$ are the mass and the radius of the star at the time when the outflow 
switches off.  We expect that $\mathcal{T}^{on}_{star}>\mathcal{T}^{off}_{star}$. If this is to hold, then:

\begin{equation}
\left(\frac{\mathcal{M}_{star}^{off}}{\mathcal{M}_{star}^{on}}\right)^{1-\zeta_{0}}
\left(\frac{\mathcal{R}^{off}_{star}}{\mathcal{R}^{on}_{star}}\right)^{3\zeta_{0}-1}<\left(\frac{9}{192}\right)^{\frac{2}{5}}=0.30.
\end{equation}

\noindent Hence, outflow activity will take place when: $(\mathcal{T}^{off}_{star}\leq\mathcal{T}_{star}\leq\mathcal{T}^{on}_{star})$. When 
$\mathcal{T}_{star}= \mathcal{T}^{off}_{star}$, we have: $\epsilon_{1}=9\mathcal{R}_{star}^{s}/192\mathcal{R}_{star}$. Using the approximate relation 
for an accreting star:  $\mathcal{R}_{star}~\sim~61\mathcal{R}_{\odot}\left(\mathcal{M}_{star}/\mathcal{M}_{\odot}\right)$, one arrives at: 
$\epsilon_{1}=3.32\times10^{6}$. This means: $(\epsilon_{1}^{on}=1)$, and: $\epsilon_{1}^{off}=3.32\times10^{6}$, where $\epsilon_{1}^{on}$ and 
$\epsilon_{1}^{off}$, are the values of $\epsilon_{1}$ when the outflow switches on and off respectively,  hence thus outflow activity will take place 
during which period when:

\begin{equation}
1\leq\epsilon_{1}< 3.32\times10^{6}.
\end{equation}

The emerging picture is that $ \mathcal{T}$ gets larger and larger as the star accretes  more and more matter until a peak moment is reached (most 
probably when the star stops growing in mass) where upon the spin begins to slow down, in which process of slowing down the inner cavity inside the lob 
of the outflow is created. This inner cavity grows bigger and bigger as the star's spin slows down, until such a time when the spatial dimensions of 
this cavity is equal to the outflow lobe itself. Once this state is attained, the outflow switches off because the growing cavity has -- eaten up from 
within, all the outflow region. 

Clearly, the above picture suggests that the spin of a star is what controls outflows, at some specific state, the outflow switch's-on; it  evolves to 
some peak spin-value; thereafter, its spin slows down. This means that during the outflow process after the begins to slow down, the star loses some 
spin angular momentum. This idea resonates with the long held suggestion discussed earlier that outflows are thought to exist as one means to tame the 
spin angular momentum of a star (see \textit{e.g.} Larson $2003b$). We will not go deeper than this in our analysis. The aim has been to show that the 
emergent picture of outflows from the ASTG is capable (in principle) to answer such questions. This means in a future study, these are the things to 
look forward to.

\subsection{Outflow Feed Region}

In the Outflow Feed Region -- \textit{i.e.} the region in figure \ref{outfdiag} described $\textrm{OEF}$ and $\textrm{OGH}$, clearly, any material that 
enters this region is going to be channeled into the Outflow Region because the repulsive radial component of the gravitational field (aided by the 
radiation field) is going to channel this matter radially outward  while the azimuthal component is going to going to channel this outward radially 
moving material toward the spin axis, hence it is expected that most of the matter will enter the Outflow Region along the the spin axis of the star. 
It is important to state that no matter the radiation from the star, there will be no reversal of in-falling matter outside the region of repulsive 
gravitation due to the radiation field of the nascent star -- we shall discuss this in \S $(8)$.

\subsection{Outflow Region}

The Outflow Region is comprised of a section of a cone ($\textrm{OAB}$ \textit{\&} $\textrm{OCD}$), the outflow loboid minus the Outflow Feed Region. In 
this Outflow Region, the gravitational force is both  radially azimuthally repulsive \textit{i.e.}, $(g_{\theta}>0)$ and $(g_{r}>0)$. This means, once 
the repulsive gravitational force is switched-on and it is in a fully fledged phase, all material found in this region is going to be channeled out of 
this region radially along with most of the matter concentrated along the spin axis. The material will be concentrated along the spin axis because the 
repulsive azimuthally gravitational component will  channel toward the spin axis. The repulsive radial component pushes the material out radially, while 
the repulsive azimuthal component of the gravitational force draws  this material close the spin axis hence the bulk of the outflow material must be 
found along the edge spin axis.

Where the cone meets the outflow loboid \textit{i.e.}, along $\textrm{AB}$ and $\textrm{CD}$, there is going to be rings. Considering the ring AB, it is 
clear that this ring (as $\textrm{CD}$) must be a shock front since  on this ring,  along the radial line $\textrm{OA}$, the in-coming material will meet the outgoing material with equation but opposite radial forces. This equal and but opposite forces must create (radially) a stationery shock. This shock is going to have a ring structure -- let us call this the \textit{Shock Ring}. As the rings $\textrm{AB}$ \textit{\&} $\textrm{CD}$,  $\textrm{EF}$ \textit{\&} $\textrm{GH}$ will be rings too, but not shock things. These rings $\textrm{EF}$ \textit{\&} $\textrm{GH}$ are the mouth of the outflow and matter enters in to the outflow region \textit{via} this opening.

\subsection{Shock Rings and Methanol Masers}

Given that $(1):$ $\textrm{AB}$ \textit{\&} $\textrm{CD}$ are shock rings, $(2):$ that methanol masers (amongst other pumping mechanisms) are  thought 
to arise in shock regions and $(3):$ the observations of Bartkiewicz \textit{et al.}  (\cite{bartkiewicz05}) where these authors discovered a ring distribution of $6.7\, \textrm{GHz}$ methanol masers, it is logical to assume that  this shock ring may well be a hub of methanol masers arising from the shock present on this ring. Recent and further work by these authors strongly suggests that a Ring \textit{of} Masers is a natural occurrence in star forming regions as (Bartkiewicz \textit{et al.} \cite{bartkiewicz09}). 

This ring distribution of masers components, they believe strongly suggests the existence of a central source -- this is the case here, the central 
source must exists and it is the forming star. They found an infrared object coinciding with the center of the ring of masers within $78\, mas$ and 
this source is cataloged in the 2MASS survey  as $\textrm{2MASS183451.56-08182114}$. They believe this is an evolving evolving protostar driving this 
masers \textit{via} circular shocks -- this is in line with the the present.  Very strongly, the Bartkiewicz Ring \textit{of} Masers suggests -- in our 
opinion that; our outflow model may very well contain an element of truth, that our model contains the possible seeds of resolution of this puzzling 
occurrence of Ring Masers.

About this shock ring; when viewed from the projection as shown in figure \ref{outfdiag}, the distance of the shock ring from the star will be:

\begin{equation}
l_{sh}=l_{max}(3)^{-\frac{3-\alpha_{\rho}}{4-2\alpha_{\rho}}}\label{shock},
\end{equation}

\noindent and the radius of this shock ring will be:

\begin{equation}
\mathcal{R}_{ring}=l_{max}(1.5)^{-\frac{3-\alpha_{\rho}}{4-2\alpha_{\rho}}}\label{ring}.
\end{equation}

\noindent Clearly, for an isolated system, depending on the orientation relative to the observer, this ring can appear as a linear structure, a circular
 or an elliptical ring. 

At present more than $500$ $6.7\, \textrm{GHz}$ methanol masers sources are known to exist (Malyshev \textit{\&} Sobolev \cite{malyshev03}; Pestalozzi 
\textit{et al.} \cite{pestalozzi05}; Xu \textit{et al.} \cite{xu03}) and are associated with a very early evolutionary phase of high mass star 
formation. The methanol maser emitting at the $6.7\,\textrm{GHz}$ frequency first discovered by \cite{menten91} is the second strongest centimeter 
masing transition of any molecule (after the $22\,\textrm{GHz}$ water transition) and is commonly found toward star formation regions. It is typically 
stronger than $12.2\,\textrm{GHz}$ methanol masers (discovered by Batrla \textit{et al.} \cite{batrla87}) observed toward the same region. Methanol 
masers have become well established tracers or sign spots of high mass star formation regions. It is thought that methanol masers occur in the very 
early stages of massive star formation. 

While methanol masers are found in regions of massive star formation, some have been found with no associated high mass star formation actively 
(see \textit{e.g.} Ellingsen \textit{et al.} \cite{ellingsen96}, Szymczak \textit{et al.} \cite{szymczak02}. Besides this non-association, some methanol
 masers are and have been observed to exist in close spatial proximity of massive stars. This has lead to the classification of methanol masers into 
Class I and Class II. Class I masers emit at the frequencies $25.0$, $44.0$, $36.0\,\textrm{GHz}$ etc while class II methanol masers emit at 
$6.7$, $12.2$, $157.0\,\textrm{GHz}$ \textit{etc} methanol masers is classified as Class II. Class I methanol masers are often observed to exist apart 
from  the continuum sources , while Class II are observed to exist very close, albeit, both classes often co-exist in the same star forming region 
inside an HII regions (\textit{e.g.} Sobolev \textit{et al.} \cite{sobolev04}). Clearly, $l_{sh}=l_{sh}(t)$ and $\mathcal{R}_{ring}=
\mathcal{R}_{ring}(t)$ and as the star evolves, $l_{sh}$ and $\mathcal{R}_{ring}$ get larger. This means in the case of young stars, if this ring is a 
hub of methanol masers, it is expected that methanol masers will be found closer to the star for young HMS and likewise, for more evolved massive stars, 
methanol masers will be found further from the nascent star. If this is correct, then it may explain the aforesaid; why Class II methanol masers are 
mostly found close to the nascent star and why Class I methanol masers are found existing further from the nascent star.

High resolution imaging of the $6.7$ and $12.2\,\textrm{GHz}$ methanol masers has found that many exhibit a simple elongated linear or curved spatial 
morphologies (Norris \textit{et al.} \cite{norris88}; Norris \textit{et al.} \cite{norris93}; Minier \textit{et al.} \cite{minier00}) and as already 
stated, depending on the orientation of the observer relative to the star forming system, the ring may appear as a linear structure. These linear 
structures have lengths of $50$ to $1300\, \textrm{AU}$. Because of this, one of the possible interpretations that has been entertained for sometime is  
that the masers originate in the circumstellar accretion disc surrounding the newly formed star (Edris \textit{et al.} \cite{edris05}) and besides this; 
because of their strong association with outflows (see \textit{e.g.} Plambeck \textit{\&} Menten \cite{plamberk90}; Kalenskii \textit{et al.} 
\cite{kalenskii92}; Bachiller \textit{et al.} \cite{bachiller95}; Johnston \textit{et al.} \cite{johnston92}), other than originating from the 
circumstellar disk, also, it has been entertained that methanol maser may originate from outflows (see \textit{e.g.} Pratap \textit{\&} Menten 
\cite{pratap92}; de Buizer \textit{et al.} \cite{debuizer00}). Clearly, the outflow origin of methanol masers resonates with the present ideas. If the 
ideas herein are correct, then, this reading would of value to researchers seeking an outflow origin of methanol masers.

Further, if viewed from the same view as in figure \ref{outfdiag}, and if as argued above that masers are found on the ring, one will expect to observe 
a linear alignment of masers above and below the the nascent star. This would explain the observed linear alignment of methanol masers  and also the 
observed linear alignment of masers above an below the IRAS source found in molecular cloud $\textrm{G69.489-0.785}$ (see Fish \cite{fish07}). Given 
Fish's observations of blue and red-shifted masers in the \textrm{ON1}-region (Fish \cite{fish07}), the suggested model of this ring of masers is 
interesting as it may offer an explanation  of this unexplained and puzzle of red and blue-shifted masers  at opposite sides of the IRAS source 
associated with \textrm{ON1}. 

\subsection{ Collimation Factor} 

We can calculate the collimation factor of the outflow since we know the extent ($l_{max}$) and the breath of the outflow which is the size of the 
shock rings \textit{i.e.}, the collimation factor could be: $q_{col}=\mathcal{R}_{ring}/l_{max}$, which can  also be written as: 

\begin{equation}
q_{col}=(1.5)^{\frac{3-\alpha_{\rho}}{4-2\alpha_{\rho}}},\label{colf}
\end{equation}

\noindent (this has been deduced from equation \ref{ring}). Now, it is believed that the most stable density profile is one with a density index 
$(\alpha_{\rho}=2)$, this means molecular clouds in a state different from this density profile will tend to it. Using this assumption, we see that  
as: $(\alpha_{\rho}\longmapsto2)$ from $(\alpha_{\rho}=0)$, \textit{i.e.} $(\alpha_{\rho}:0\longmapsto2)$,  then we will have:  
$(q_{col}\longmapsto\infty)$. For this setting, generally: $(q_{col}>1)$. We also realize that now as: $(q_{col}\longmapsto\infty)$,  when: 
$(\alpha_{\rho}: 3\longmapsto 2)$, then: $(q_{col}>1)$, and if:  $(\alpha_{\rho}: 0\longmapsto 2)$. For this setting, generally: $(q_{col}\geq1.36)$. 
This means we are going to have two categories of collimation factor value i.e. $(1<q_{col}<1.36)$ for $(\alpha_{\rho}: 0\longmapsto 2)$ and 
$(q_{col}\geq1.36)$ for  $(\alpha_{\rho}: 3\longmapsto 2)$.  Because of projection effects, it is very difficult  to measure the true collimation 
factor.

Also, because of projection effects, the collimation factor that we measure in real life is not the actual collimation factor but the projected 
collimation factor. If we know the actual collimation factor, we will be able to know the density index since from (\ref{colf}) we can deduce that: 

\begin{equation}
\alpha_{\rho}=2-\left(\frac{\log q_{col}^{2}}{\log 1.5}-1\right)^{-1}=\frac{\log (q_{col}^{4}/8)}{\log (1.5)}.\label{colf2}
\end{equation} 

\noindent LMSs are known to have relatively low outflow collimation factors ($q_{col}<2$) while HMSs have significantly high outflow collimation factors ($2<q_{col}<10$), sometimes reach $q_{col}\sim20$. From (\ref{colf2}) the aforesaid implies, assuming these collimation factors are a good representation of the 
real collimation factor, that LMSs cores have density index $\alpha_{\rho}=1.56$ and HMS cores have density index $\alpha_{\rho}=1.98$. This is not 
unreasonable but very much expected. The fact that for HMS forming cores, we have $\alpha_{\rho}=1.98$ and for LMS forming cores we have 
$\alpha_{\rho}=1.56$, means HMS cores are much more dense compared to LMS forming cores.

\section{Radiation Problem\label{srad}}

While the main thrust and focus of this reading is not on the \textit{Radiation Problem} associated with massive stars, but on the polar repulsive 
gravitational field and its possible association with the observed bipolar molecular outflows, we find that the ASTG affords us a window of opportunity 
to visit this problem. This so-called radiation problem associated with massive stars has been well articulated in Paper III. There is no need for us to go through the details of this same problem here but we shall direct the reader to Paper III for an 
exposition of the radiation problem. In the subsequent paragraphs, we shall -- for the sack of achieving a smooth continuous reading; present  the 
findings of Paper III in nutshell.

In general, a massive star is defined to be one with mass greater than $\sim8-10\mathcal{M}_{\odot}$ and central to the on-going debate on how these 
objects [massive stars] come into being is this so-called radiation problem. For nearly forty years, it has been argued that  the radiation field 
emanating from massive stars  is high enough to cause a  global reversal of direct radial in-fall of  material onto the nascent star. In Paper III, it 
is argued that only in the case of a non-spinning isolated star does the gravitational field of the nascent star overcome the radiation field. An 
isolated non-spinning star is a non-spinning star without any circumstellar material around it, and the gravitational field beyond its surface is 
described exactly by Newton's inverse square law. The supposed fact that massive stars have a gravitational field that is much stronger than their 
radiation field is drawn from the analysis  of a non-spinning isolated massive star. In this case, the gravitational field is (correctly) much stronger 
than the radiation field. This conclusion has been erroneously extended to the case of non-spinning massive stars enshrouded in gas and dust. 

It is argued there, in Paper III, that, for the case of a non-spinning gravitating body where the circumstellar material is taken into consideration, 
that at $\sim8-10\mathcal{M}_{\odot}$, the radiation field will not reverse the radial in-fall of matter, but rather a stalemate between the radiation 
and gravitational field will be achieved,  \textit{i.e.} in-fall is halted but not reversed. Any further mass growth is stymied and the star's mass 
stays constant at $\sim8-10\mathcal{M}_{\odot}$. This picture is very different from the common picture that is projected and accepted in the wider 
literature where at $\sim8-10\mathcal{M}_{\odot}$, all the circumstellar material, from the surface of the star right up to the edge of the molecular 
core, is expected to be swept away by the all-marauding and pillaging radiation field. There in Paper III, it is argued that massive stars should be 
able to start their normal stellar processes if the molecular core from which they form has some rotation, because a rotating core exhibits an ASGF 
which causes there to be an accretion disk and along this  disk the radiation is not powerful enough to pillage the in-falling material. We show here 
that in the region: ($\theta:\,[125.3 <\theta<54.7]$ \textit{\&} $[234.7<\theta<305.3]$), around a spinning star the gravitational field in the face of 
the radiation field,  will never be overcome by the radiation field  hence in-fall reversal does not take place in this region and this region is the 
region \textit{via} which the nascent massive star forms once the repulsive outflow field and the star's mass has surpassed the critical 
$8-10\,\mathcal{M}_{\odot}$. Reiterating, in this region \textit{i.e.} ($\theta:\,[125.3 <\theta<54.7]$ \textit{\&} $[234.7<\theta<305.3]$), infall is 
never halted but continues unaborted and unabated.

There are three cases of an embedded spinning nascent star $(1):$ Where the nascent star is spinning and the circumstellar material is not spinning or 
where the spin of the circumstellar material is so small compared to the star so much that the circumstellar material can be considered to be not 
spinning. $(2):$ Where  the nascent star is spinning independently of the  circumstellar material which is itself  spinning. $(3):$ Where the nascent 
star is co-spinning or co-rotating with the circumstellar material. It should suffice to consider one case because the procedure to show that in the 
region: ($\theta:\,[305.3<\theta<54.7]$ \textit{\&} $[234.7<\theta<125.3]$), infall is never halted but continues unaborted and unabated, is the same. 
Of the three cases stated, the most likely scenario in \textit{Nature} is the second case \textit{i.e.}, where  the nascent star is spinning 
independently of the  circumstellar material which is itself  spinning. We shall consider this case.

The ASGP for the case of a star that is spinning independently of its core has be argued to be given by (\ref{pot-star-csl}) and in the face of 
radiation field, the resultant radial component of the gravitational field intensity is given by:

\begin{equation}
g_{r}(r,\theta)=-\frac{G\mathcal{M}_{0}^{eff}}{r^{2}}\left[1- \frac{\kappa \mathcal{L}_{star}}{4\pi G\mathcal{M}_{0}^{eff}c}+
\frac{2\lambda_{1}^{star}\gamma_{1}G\mathcal{M}_{1}^{eff}\cos\theta}{rc^{2}}+3\lambda_{2}^{star}\gamma_{2}
\left(\frac{G\mathcal{M}_{2}^{eff}}{ rc^{2}}\right)^{2}\left(\frac{3\cos^{2}\theta-1}{2}\right)\right].\label{radcomp}
\end{equation}

\noindent For the radiation component to be attractive, we must have: $[g_{r}(r,\theta)<0]$, and for this to be so, the term in the square brackets must 
be greater than zero, this implies:

\begin{equation}
\left[1-\frac{\kappa \mathcal{L}_{star}}{4\pi G\mathcal{M}_{0}^{eff}c}\right]r^{2}+
\left[\frac{2\lambda_{1}^{star}\gamma_{1}G\mathcal{M}_{1}^{eff}\cos\theta}{c^{2}}\right]r+
\left[3\lambda_{2}^{star}\gamma_{2}\left(\frac{G\mathcal{M}_{2}^{eff}}{ c^{2}}\right)^{2}\left(\frac{3\cos^{2}\theta-1}{2}\right)\right]>0.
\label{radcomp2}
\end{equation}

\noindent This inequality is quadratic in $r$ and can be written as: $(Ar^{2}+Br+C>0)$, where: $A,B,$ and $C$, can easily be obtained by making a 
comparison. Since\footnote{In Paper I, we  argued that, $r$ can take both negative and positive values, and further argued that the set up of the 
coordinate system of the ASGF is such that [$r>0$ \& $\cos\theta>0$] and [$r<0$ $\cos\theta<0$], hence $r\cos\theta>0$, which implies $(Br>0)$.}: 
$(Br>0)$, for: $(Ar^{2}+Br+C>0)$, to hold absolutely, we must have: $(Ar^{2}>0\Rightarrow A>0)$ and $(C>0)$. The condition: $(A>0)$, implies:

\begin{equation}
\mathcal{M}(r)>\frac{\kappa_{eff} \mathcal{L}_{star}}{4\pi Gc}.\label{cavity}
\end{equation}

\noindent To arrive at the above one must remember that: $\mathcal{M}_{0}^{eff}=\mathcal{M}_{star}+\mathcal{M}_{csl}(r)=\mathcal{M}(r)$. As shown in 
Paper II (see equation $5$ \textit{\&} \S $5$ of Paper II), the condition (\ref{cavity}) for: $\mathcal{M}_{star}>8-10\,\mathcal{M}_{\odot}$, leads to 
the formation of a cavity inside the star forming core. In this cavity, the radiation field in powerful enough to halt infall reversal but outside of 
it, it is not.

Now, for the condition: $(C>0)$, to hold (remember $\lambda^{star}_{2}<0$), this means: $(3\cos^{2}\theta-1<0)$, hence: 
($\theta:\,[125.3 <\theta<54.7]$ \textit{\&} $[234.7<\theta<305.3]$). The result just obtained invariably means inside the cavity created by the 
radiation field, the region: ($\theta:\,[125.3 <\theta<54.7]$ \textit{\&} $[234.7<\theta<305.3]$) will have an attractive gravitational field, hence 
matter will still be able to fall onto the nascent star \textit{via} this region and this in-falling of matter is completely independent of the opacity 
of the material of the core! Hence we expect spinning massive stars to face no radiation problem at all. Clearly, if: $(\lambda_{2}>0)$, then in the 
region: ($\theta:\,[125.3 <\theta<54.7]$ \textit{\&} $[234.7<\theta<305.3]$), the gravitational field was going to cause in-fall reversal in the cavity 
hence disallowing for the star to continues is accretion. This obviously would have been at odds with experience hence thus we have the strongest 
reason for setting: $(\lambda_{2}>0)$, otherwise the ASTG would be seriously at odds with physical and natural reality as we know it. Beside, the 
condition $(\lambda_{2}>0)$ is supported by the solar data (see Paper I). The fact that in the region; ($\theta:\,[125.3 <\theta<54.7]$ \textit{\&} 
$[234.7<\theta<305.3]$), is a region of attractive gravitation, it is clearly that the ASGF will form a disk around the nascent star.  Although no 
detailed study of accretion disks has been made (Brogen \textit{et al.} \cite{brogen07}; Araya \textit{et al.} \cite{araya08}) and this being due 
technological challenges in obtaining must higher resolution observations on the  scale of these accretion disks, it has long been thought that the 
accretion disk is a means by which accretion of matter on the nascent stars continues soon after radiation has (significantly) sounded her  presence on 
the star formation podium (see \textit{e.g.} Chini \textit{et al.} \cite{chini04}; Beltr$\acute{\textrm{a}}$n \textit{et al.} \cite{beltran04}). If our 
investigation prove correct, as we believe they will, then, researchers have been right to think that that accretion disk serves a platform for further 
accretion of mass by the nascent star.

\section{Discussion and Conclusion}

This reading should be taken more as a genesis that lays down the mathematical foundations  that seek to lead to the resolution of the problem of 
outflows, \textit{vis}, what their origin is. Also, we should say that, if this reading is anything go by \textit{i.e.}, if it proves itself to have a 
real direct correspondence with the experience of physical reality, then not only have we laid down the mathematical foundations that may lead to the 
understanding of outflows; but we have laid a three fold foundation that could lead to the resolution of three problems, and these problems are:

\begin{quote}
\begin{enumerate}
\renewcommand{\theenumi}{(\arabic{enumi})}
\item The {Origins and Nature of Outflows}

\item The {Radiation Problem} thought to exist for HMS.

\item The {Origin of Linear \textit{\&} Ring Structures of Methanol Masers}.

\end{enumerate}
\end{quote}

\noindent All this we have arrived at after the consideration of the azimuthal symmetry arising from the spin of a gravitating body. This symmetry has 
been applied to the gravitational field and where upon we have come up with the ASTG. In Paper I, we did show that the ASTG can explain the perihelion 
shift of planets in the solar system and therein, the ASTG as it lays there, suffers the setback that the ``constants'' $\lambda_{\ell}$ are unknown. We 
have gone so far in the present as to suggest a way to solve this problem but this suggestion is subject to revision pending any new data. 

It should be said that, to the best of what we can remember ever-since we learnt that the force of gravity is what causes an apple to fall to the ground 
and that the very same force causes the moon and the planets to stay in their orbs; we have never really convinced of gravitation as being a repulsive 
force, let alone that it possibly can have anything to do with the power behind outflows.  Just as anyone would find these ideas in violation of their 
intuition, we find our-self in the same bracket. But one thing is clear, the picture emerging from the mathematics thereof, is hard to dismiss. It calls 
one to make a closer look at the what the Poisson equation is ``saying to us''.

In closing, allow me to say that as things stand in the present -- while we firmly believe we have discovered something worthwhile; it is difficult to 
make any bold conclusions. Perhaps we should only mention that work has began on a numerical model of outflows based on what we have discovered herein. 
Only then -- we believe; it will  be possible to make any bold conclusions.

\section*{ Acknowledgments}

I am grateful to my brother George and his wife Samantha for their kind hospitality they offered while working on this reading and to Mr. Isak D. Davids 
\textit{\&} Ms. M. Christina Eddington for proof reading the grammar and spelling and Mr. M. Donald Ngobeni for the magnanimous support.

\label{lastpage}
\end{document}